\documentclass[12pt]{iopart}
\usepackage[english]{babel}
\usepackage[latin9]{inputenc}
\usepackage{graphicx}
\usepackage{multirow}
\usepackage[usenames,dvipsnames]{xcolor}
\usepackage{iopams}
\usepackage{rotating}
\makeatletter
\let\@fnsymbol\@arabic
\makeatother
\definecolor{oceanboatblue}{rgb}{0.0, 0.47, 0.75}
\definecolor{orange}{rgb}{1,0.5,0}
\definecolor{goodgreen}{rgb}{0.1,0.5,0}
\definecolor{goodred}{rgb}{0.7,0,0}
\definecolor{DarkGreen}{rgb}{0.12, 0.3, 0.17}
\definecolor{calpolypomonagreen}{rgb}{0.0, 0.2, 0.13}
\usepackage[colorlinks,urlcolor=goodgreen,citecolor=blue,linkcolor=goodred]{hyperref}
\usepackage{braket}
\usepackage{cleveref}
\usepackage{lineno}
\setpagewiselinenumbers
\modulolinenumbers[5]
\usepackage{tikz}
\usepackage{tocvsec2}
\usepackage{scrextend}
\usepackage{braket}
\usepackage[normalem]{ulem}

\makeatletter

\begin{document}
\title[Energy density as a  probe of band representations in PhCs]{%Topological Classification of Photonic Crystals based on the Local Density of States \\ or \\ Real-Space Signatures of Topological Quantum Chemistry of Light via the Local Density of States of Photonic Crystals \\ or \\
Energy density as a  probe of band representations in photonic crystals}

%Alternative title if we conclude that we cannot characterize fragile topology from the LDOS
%\title{Real space signatures of TQC of light based on the Local Density of States of photonic crystals}

\author{M.~Blanco~de~Paz$^{1,3,\parallel}$, M.~A.~J.~Herrera$^{2,1}$, P.~Arroyo~Huidobro$^3$, H.~Alaeian$^4$, M.~G.~Vergniory$^{1,5}$ B.~Bradlyn$^6$, G.~Giedke$^{1,7}$, A.~Garc\'{i}a-Etxarri$^{1,7,\dagger}$, D.~Bercioux$^{1,7,\perp}$}

\address{$^1$Donostia International Physics Center, 20018 Donostia-San Sebasti\'an, Spain \\ $^2$Centro de F\'isica de Materiales (CFM-MPC) Centro Mixto CSIC-UPV/EHU, 20018 Donostia-San Sebasti\'an, Basque Country, Spain\\ 
$^3$Instituto de Telecomunica\c c\~oes, Instituto Superior Tecnico-University of Lisbon, Avenida Rovisco Pais 1, Lisboa, 1049-001 Portugal \\
$^4$Elmore Family School of Electrical and Computer Engineering, Department of Physics and Astronomy,
Purdue Quantum Science and Engineering Institute, Purdue University, West Lafayette, IN 47907, USA \\
$^5$Max Planck Institute for Chemical Physics of Solids, Dresden D-01187, Germany\\
$^6$Department of Physics and Institute for Condensed Matter Theory, University of Illinois at Urbana-Champaign, Urbana, IL, 61801-3080, USA \\
$^7$IKERBASQUE, Basque Foundation for Science, Euskadi Plaza, 5, 48009 Bilbao, Spain}
%%%%%%%
\ead{$^\parallel$maria.depaz@dipc.org,$^\dagger$aitzolgarcia@dipc.org,$^\perp$dario.bercioux@dipc.org}
\vspace{10pt}
\begin{indented}
\item[]\today
\end{indented}

\begin{abstract}
Topological Quantum Chemistry (TQC) has recently emerged as a instrumental tool to characterize the topological nature of both fermionic and bosonic band structures. TQC is based on the study of band representations and the localization of maximally localized Wannier functions. In this article, we study various two-dimensional photonic crystal structures analyzing their topological character through a combined study of TQC, their Wilson-loop spectra and the electromagnetic energy density. Our study demonstrates that the analysis of the spatial localization of the energy density complements the study of the topological properties in terms of the spectrum of the Wilson-loop operator and TQC.
\end{abstract}

\maketitle
\section{Introduction}

Photonic crystals (PhC) are systems which control the propagation of light via a periodic modulation of the system's refractive index~\cite{photoniccrystals}. This refractive index modulation can be implemented in one, two, or three dimensions.
Two-dimensional (2D) PhCs are of particular interest since the propagation of light can be decomposed into two orthogonal set of solutions, the transverse magnetic (TM) and transverse electric (TE) modes. In such systems Maxwell's equations can be recast in a scalar form for each of these sets of modes. This reformulation allows for a straightforward analogy between the eigenvalue equation for the PhC and the 2D Schr\"odinger equation in the presence of a periodic potential~\cite{sakoda,tutorial}. 

In the past few years, there has been a growing interest in topological phases of matter, especially for the case of electronic systems~\cite{bernevigbook}. In a topological system, protected states emerge on the boundary while the bulk remains gapped~\cite{hasan2010colloquium}. This property can be exploited for possible technological applications, spanning from spintronics~\cite{He_2021,Brahlek_2020} to quantum computation~\cite{Das_Sarma_2006,Lahtinen_2017}. Topological phenomena are rooted in the wave nature of electrons; thus, they are not restricted to fermionic systems but can also appear in bosonic ones. In this respect, we can find topological systems for photons~\cite{haldane2008possible,Lu_2014}, acoustic phonons~\cite{Yang_2015} and even classical wave systems~\cite{McLeish_2019,ozawa2018topological,ma2019topological,rider2019perspective,kim2020recent}.
Photonic crystals have emerged in the past years as a versatile platform for investigating topological properties~\cite{wang2009observation}. This stems from the fact that in linear dielectrics, photons are truly non-interacting. Additionally, PhCs can be cheaply and easily engineered with almost any desirable lattice structure~\cite{photoniccrystals}. In this respect, PhCs can be considered an excellent analogue quantum simulator for interesting non-interacting problems in quantum chemistry, quantum biology and solid-state physics~\cite{Aspuru_Guzik_2012}. Furthermore, when considering non-linear dielectrics, it is possible to engineer analogue quantum simulators for interacting systems as well~\cite{Hartmann_2016}. Nowadays, there is a considerable effort to identity  PhC systems with topological properties  utilizing  novel tools such as machine learning~\cite{Peano_2021,Mertz_2021}.

In analogy to condensed-matter systems, both the density of states (DOS) and the local DOS (LDOS) are helpful instruments for investigating the spectral properties of PhC systems. Specifically, the LDOS helps to understand how a classical dipole or a quantum emitter couples to the electromagnetic (EM) field in PhCs~\cite{McPhedran_2004}. However, as we will show in this work, it is not straightforward to extract from these two quantities all the necessary information to classify the topological properties of a PhC. 
Recently, there has been some effort to understand the coupling between quantum emitters to a topological PhC ~\cite{Bello2019,perczel2020topological,navarro2021photon}. The major finding is that the spontaneous decay of the quantum emitter can be strongly modified when coupled to a topological band instead of a trivial one. Further, the topological band can modify the mutual interaction among quantum emitters, and new photon-mediated interactions can be designed~\cite{Bello2019}. Some more recent work tried to connect the EM~energy density | the energy integral of the LDOS | to the effects of fractionalization that are expected in some topological systems~\cite{Shuwai_2022}.

In this manuscript, we combine the theory of topological quantum chemistry (TQC)~\cite{NaturePaper}, the Wilson-loop (WL) spectrum~\cite{Alexandradinata14} and the EM energy density to analyze the topological character of various PhCs based on the honeycomb lattice. In the following, we will focus on 2D PhCs and we will restrict the analysis of the EM modes in the photonic crystal system to TM modes.
Specifically, we consider first the breathing honeycomb PhC introduced in Ref.~\cite{wuandhu}. Contrary to the original claim, we show that this model does not present any $\mathbb{Z}_2$ topological phase. However, we confirm that this PhC presents two trivial phases characterized by bands induced from Wannier modes localized at different (maximal) Wyckoff positions~\cite{Matt}. Here, we further address this point by showing that the EM energy density---while for the bands considered is always maximal areas with a larger dielectric constant---has peaks that are shifted towards the Wyckoff position at which the Wannier functions are centered. Secondly, we consider the combination of two breathing honeycomb PhCs, that we name the nested breathing honeycomb PhC. This structure presents various phases, including a non-trivial topological one allowed by crystalline symmetry, specifically a \emph{fragile} topological phase~\cite{Cano_2017,Po_2018,ahn2018higher,Slager2018,FragileLight}. The various phases are obtained by varying the permittivities (\emph{i.e.}, changing the dielectric materials composing the PhC) or the geometry (distances and radii of the rods). For both PhCs, we show how an analysis of the spatial localization of the EM field based on the EM energy density complements the study of the topological properties in terms of the spectrum of the WL operator and TQC. 

The article is structured as follows: in Sec.~\ref{WLSec} we give an introduction to the concept of the Wilson-loop operator and to the maximally localized Wannier functions. In Sec.~\ref{TQCSec} we briefly introduce TQC and how it can be used to study topology of PhCs. We complete the methodological overview with Sec.~\ref{SecLocal}, where we introduce the dyadic Green's function for a PhC and the definitions of DOS, LDOS, and the EM energy density. Finally, we present several examples of PhC with different topological character: in Sec.~\ref{ex_One} we consider the case of a breathing honeycomb lattice that presents trivial topology. In Sec.~\ref{ex_Two}, we present the case of a nested breathing honeycomb lattice that presents several phases including a topological fragile one. We conclude the manuscript with a discussion of results  presented in Sec.~\ref{Outlook}. We also include two technical appendices. In~\ref{app_one}, we review how to evaluate the spectrum of the Wilson loop when dealing with a discretized first Brillouin zone. In~\ref{app_two} we show how the total EM energy density introduced in Sec.~\ref{SecLocal} can be obtained starting from the EM Poynting vector.

\section{Wilson loops and Maximally-Localized Wannier Functions}\label{WLSec}

The spectrum of the Wilson-loop (WL) operator~\cite{Alexandradinata14,Alexandradinata_2014} gives a powerful and universal method for characterizing the topological phases of periodic wave systems. In the following, we present this method and its implications for PhCs. 

We begin by defining the non-Abelian Berry connection $\mathbf{A}_{mn}$ for a set of $N$ ``occupied''\footnote{Although we consider bosonic PhC systems, we will use the terminology from condensed matter physics and refer to the set of bands below the band gap of interest as occupied.} and possibly degenerate bands~\cite{Alexandradinata14,Alexandradinata_2014,Neupert2018}
%
%
%%%%%%%%%%%%
\begin{equation}
\mathbf{A}_{mn}(\mathbf{k})= \mathrm{i} \langle u_{m\mathbf{k}}|\nabla_{\mathbf{k}}|u_{n\mathbf{k}}\rangle, ~~~m,n=1,\ldots N \, ,
\label{eq:Berry_conn}
\end{equation}
%%%%%%%%%%%%
%
%
where $m,\,n$ are the band indices, $\mathbf{k}$ is the crystal momentum defined inside the first Brillouin zone (BZ), $|u_{n\mathbf{k}}\rangle$ is the periodic part of the Bloch eigenfunctions defined inside the real-space unit cell (UC), and $\nabla_{\mathbf{k}}$ is the gradient with respect to crystal momentum $\mathbf{k}$.
The (abelian) Berry phase is defined as the line integral of the Berry connection along a closed path in momentum space, traced over the band indices:
%
%
%
%%%%%%%%%%%%%%%%
\begin{equation}
\phi(\ell)=\oint_\ell \mathrm{Tr }\,\mathbf{A}_{mn}(\mathbf{k})\cdot d\boldsymbol{\ell}\qquad \mathrm{mod }2\pi\, .
\label{eq:BP}
\end{equation}
%%%%%%%%%%%%%%%
%
%
For a contractible path, we can use Stokes's theorem to transform the line integral into a surface integral, which allows to define the Berry curvature:
%
%
%%%%%%%%%%%%%%%%
\begin{equation}
\phi(\ell)=\oint_\ell  \mathrm{Tr}\,\mathbf{A}_{mn}(\mathbf{k})\cdot d\boldsymbol{\ell}=\int_\mathcal{S}  \mathrm{Tr }~{\bOmega}_{mn}(\mathbf{k})\cdot d^2\boldsymbol{S}\, ,
\label{eq:BC}
\end{equation}
%%%%%%%%%%%%%%
%
%
where ${\bOmega}_{m,n}(\mathbf{k})=\nabla_{\mathbf{k}}\times\mathbf{A}_{mn}(\mathbf{k})$ is the Berry curvature, $\boldsymbol{S}$ is the surface defined by the closed loop $\ell\equiv\partial\mathcal{S}$, and $d^2\mathbf{S}$ is the positively-oriented surface  differential. When the integration of the Berry curvature is done over the entire first BZ, the result is quantized in multiples of $2\pi$~\cite{vanderbilt2018}. This allows to define a topological invariant known as the Chern number:
%
%
%%%%%%%%%%%%%%
\begin{equation}
C=\frac{1}{2\pi}\int_\mathrm{BZ}  \mathrm{Tr }~\boldsymbol{\Omega}_{mn}(\mathbf{k})\cdot d^2\boldsymbol{S} \, .
\label{eq:Chern}
\end{equation}
%%%%%%%%%%%%%%
%
%
The Chern number can take non-zero integer values only when time-reversal symmetry (TRS) is broken~\cite{bernevigbook}.

If instead of tracing Eq.~(\ref{eq:BP}), we consider the path-ordered exponential, we obtain a unitary matrix defining the WL operator~\cite{Alexandradinata14,Alexandradinata_2014,Neupert2018}, \emph{i.e.},
%
%
%%%%%%%%%%%%
\begin{equation}
{\cal{W}}_{mn}(\ell)= {\cal{P}}\mathrm{e}^{-\mathrm{i}\oint_\ell  {\mathbf{A}}_{mn}(\mathbf{k})\cdot d\boldsymbol{\ell}}, 
\label{eq:WL}
\end{equation}
%%%%%%%%%%%%
%
%
where $\cal{P}$ denotes the path ordering operator of the exponential. For a single isolated band, we have $m=n$ in the Berry connection~(\ref{eq:Berry_conn}), and the path ordering operator becomes trivial due to the Abelian character of the single-band Berry connection. We can establish a direct connection between the spectrum of the Wilson loop evaluated in the first BZ and the Chern number: the slope will tell us about the sign of the Chern number, whereas the number of windings  will tell us its absolute value. In \ref{app_one}, we explain how to evaluate the spectrum of the Wilson-loop operator when dealing with a discretized version of the first BZ.

There is a deep connection between the WL operator and Wannier functions. The Wannier functions are expressed as a Fourier transformation of the Bloch modes:
%
%
%%%%%%%%%%%%%%%%%
\begin{equation}\label{BlochToWannier}
    w_{i\mathbf{R}}(\mathbf{r})\propto \int_\mathrm{BZ} d\mathbf{k} \mathrm{e}^{-\mathrm{i}\mathbf{k}\cdot\mathbf{R}}\sum_j U_{ij}^{\mathbf{k}}\psi_{j\mathbf{k}}(\mathbf{r})\, ,
\end{equation}
%%%%%%%%%%%%%%%%
%
%
where $\mathbf{R}$ is a lattice vector and $U_{ij}^{\mathbf{k}}$ is a unitary matrix of dimension $N$ periodic in $\mathbf{k}$. This matrix represents the mixing of the Bloch modes in reciprocal space~\cite{Marzari2012} which are defined as $\psi_{j\mathbf{k}}(\mathbf{r})=\mathrm{e}^{\mathrm{i} \mathbf{k}\cdot\mathbf{r}}u_{j\mathbf{k}}(\mathbf{r})$. The proportionality  in Eq.~(\ref{BlochToWannier}) depends on the real space dimensionality.

For the \emph{maximally-localized} Wannier functions (MLWFs), the mixing matrix $U_{i,j}^{\mathbf{k}}$ is chosen to minimize the delocalization of the wave-function in real space according to the sum of the quadratic spreads of the Wannier functions~\cite{Marzari2012,marzari1997maximally}. The sum of the phases of the eigenvalues of the WL operator taken over a straight-line path through the first BZ corresponds to the expectation value, modulo $2\pi$, of the projected position operator evaluated over the MLWFs~\cite{Alexandradinata_2014,Neupert2018}.

For a topologically trivial system, the Wilson loop eigenvalues are adiabatically deformable to a constant value. This implies immediately that the WL spectrum does not wind, such that the Chern number must be zero. In this case, the MLWFs are exponentially localized with a well-defined position. Since crystal symmetries constrain the centers of MLWFs, we can distinguish between different topologically trivial phases by looking at where in the UC the MLWFs are localized. In many PhC applications, the MLWFs for a given set of bands are localized at the center of the UC | in this work we will refer to this as the trivial case for reference. Another possible topologically trivial phase is the so-called \emph{obstructed atomic limit} (OAL), which presents non-winding but displaced values of the WL eigenvalues. For this case, the MLWFs are still exponentially localized, but in this case their positions are located between consecutive UCs instead of at the origin. In electronic  systems, the position of the  MLFWs relative to the atomic positions relates to the bulk electronic multipole moments, which can be quantized in the presence of crystal symmetries~\cite{benalcazar2019quantization,BenalcazarScience,benalcazar2017quadPRB}.

On the contrary, for a system presenting a nontrivial topological phase~\cite{Shen_2013,Moore_2017}, the eigenvalues may wind as a function of the momentum, meaning that the WL spectrum presents a variation of $2\pi n$ with $n\in\mathbb{Z}$ along the first BZ. In this case, MLWFs respecting the crystal symmetries have a localization that is no longer exponential within the UC, but polynomial. 

Lastly, we distinguish between ``strong'' topology and the recently-discovered topological phases  exhibiting  \emph{fragile} topology~\cite{Cano_2017,Po_2018}, that has been also generalized to the case of PhCs~\cite{tutorial,FragileLight}. Fragile topological phases protected by $C_2$ and TRS display WL spectra composed of two opposite windings which indicates that, although the total Chern number is equal to zero, the MLWFs are delocalized within the UC. These systems presents spectral and topological features similar to those of $\mathbb{Z}_2$ insulators~\cite{Shen_2013}. For both phases, TRS ensures that the Chern number is zero. The WL spectrum for both cases is identical, but behaves very differently when a new set of trivial bands is added to the non-Abelian Berry connection~(\ref{eq:Berry_conn}). For topologically fragile systems, the windings are transformed into a trivial WL spectrum similar to the OAL phase when extra bands are added. On the other hand, for the strong $\mathbb{Z}_2$ phase, the winding of the WL spectrum is preserved even after the addition of trivial bands.

We can adapt all the concepts that we have introduced to PhC systems. However, the interpretation is different since we are not dealing with electron charges: we associate the photonic MLWFs to the EM energy density | see  Sec.~\ref{SecLocal} and~\ref{app_two}. A second significant difference is that in electronic systems the atoms sit in specific locations called Wyckoff positions (WPs)~\cite{Dresselhaus_2007,Cano_2021}. In contrast, for a PhC, we introduce the concept of  \emph{photonic particles}: we define this as the collection of dielectric objects in the UC~\cite{tutorial}. The dielectric objects in the photonic particle can be placed anywhere within the UC, but their \emph{center of mass} will still be located at a WP. In electronic systems we know that the basis states for a band structure are composed of atomic orbitals localized on the atomic positions. In contrast, it is only for the lowest frequency bands that we expect photonic MLWFs to be localized on the photonic particles (corresponding to the tendency of high-dielectric materials to trap long-wavelength EM modes).

In summary, Berry-connection-related quantities, such as Berry phase, Chern number and WL spectrum, are powerful tools to extract information about topological properties. In this work, we will be focused on the extent to which the EM energy density distribution for a set of bands can be understood from the WL and photonic MLWFs.

\section{Topological Quantum Chemistry of light\label{TQCSec}}

In this Section, we briefly describe how to apply the theory of TQC to PhCs. The constituents of atomic or molecular crystals are always placed at the WPs where the basis atomic orbitals are placed as well, provided they are exponentially localized. Topological quantum chemistry analyzes the irreducible representations (\emph{irreps}) at every point in the Brillouin zone induced from these atomic orbitals. On the contrary, for PhCs, the location of the dielectric within the UC is not directly connected to the position of any basis functions for the photonic band structure. 
Therefore, the most practical approach to analyze PhCs starts from reciprocal space. 

Once we have identified the space group of the lattice, we compute the fields at each high-symmetry point, taking into account that each eigenmode transforms under a representation of the little group determined by its crystal momentum. Using the Mathematica package GT-Pack~\cite{gtpack1,gtpack2},  we extract the corresponding collection of \emph{irreps} under which the Bloch modes at high-symmetry points transform. Then, we can seek to identify the band representation (BR) under which different groups of bands transform. The BRs of a space group can (in most cases\cite{cano2022topology}) be identified by the collection of \emph{irreps} for sets of bands separated by gaps. The set of irreps for a collection of bands must be compared with the elementary band representation (EBR) of the space group that is available on the Bilbao Crystallographic Server~\cite{BCS}. If the set of irreps of the collection can be expressed as a sum of EBR with positive integer coefficients, then the bands can be trivial. If the set of irreps cannot be expressed as a sum of EBRs with positive integer coefficients, then the bands are topological, meaning that the Wannier functions are not exponentially localized. Finally, if in such a decomposition, all coefficients are integers, but some necessarily negative, then the set of irreps indicate that this collection of band displays a \emph{fragile} topology. 

\section{Local, total density of states and EM energy density}\label{SecLocal}

To conclude, we present some local observable that will allow to visualize the effects of the topological markers introduced in the previous two sections. Specifically, in this Section we introduce the concept of LDOS, DOS and EM energy density of PhCs. 

To proceed, we start from the dyadic Green's function for a PhC~\cite{Sakoda_1996,Glauber_1991,Dowling_1992,Sakoda_2005}
%
%
%%%%%%%%%%%%%%
\begin{eqnarray}\label{greens}
\hspace{-2.2cm}  \mathbf{G}(\mathbf{r},\mathbf{r'};\omega) & = \int_{-\infty}^\infty dt ~ {\mathbf{G}}(\mathbf{r},\mathbf{r'};t) \mathrm{e}^{\mathrm{i}\omega t}\nonumber \\
& = \frac{c^2}{V}\int_{\mathrm{BZ}}\!\!\!\! d\mathbf{k}  \sum_{n}\sqrt{\varepsilon(\mathbf{r})\varepsilon(\mathbf{r}')} \!\left[\frac{\mathbf{E}^{(\mathrm{T})}_{\mathbf{k}n}(\mathbf{r}) \otimes \mathbf{E}^{(\mathrm{T})*}_{\mathbf{k}n}(\mathbf{r'})}{ \omega^2-\omega_{\mathbf{k}n}^2} %\right.\nonumber\\
%& \left.
+ \frac{\mathbf{E}^{(\mathrm{L})}_{\mathbf{k}n}(\mathbf{r}) \otimes \mathbf{E}^{(\mathrm{L})*}_{\mathbf{k}n}(\mathbf{r'})}{\omega^2}\right]\!,
\end{eqnarray}
%%%%%%%%%%%%%%
%
%
where we have quasi-transversal (T) and quasi-longitudinal (L) components of the electric field\footnote{We call these quasi-transverse to distinguish them from the transverse modes we obtain in the case of a homogeneous system $\varepsilon(\mathbf{r})=\varepsilon$~\cite{Sakoda_1996,Sakoda_2005}. We will omit the T and the L for simplifying the notation from here onward.} $\mathbf{E}^{(\mathrm{T/L})}_{\mathbf{k}n}(\mathbf{r},t)$ for the $n$-th band with eigenfrequency $\omega_{\mathbf{k}n}$ and momentum $\mathbf{k}$. The integration in Eq.~(\ref{greens}) is performed over the first BZ. In the following, we focus on 2D PhCs and we will work in the basis of TM modes. These PhC modes have a non-zero electric field component only along the longitudinal axis of the rods, $E_z$. As a consequence, the last term in Eq.~(\ref{greens}) is zero, the only non-zero entry of $\mathbf{G}(\mathbf{r},\mathbf{r}',t)$ is $G_{zz}$, and the volume $V$ coincides with the area $S$ of the UC in real space. The denominator of Eq.~(\ref{greens}) should be understood to have $\omega\rightarrow\omega+\mathrm{i}0^+$ to ensure causality when the frequency of the dipole coincides with the eigenfrequency, \emph{i.e.}, $\omega = \omega_{\mathbf{k}n}$. The dyadic Green's function describes the electromagnetic field at the position $\mathbf{r}$ generated by a point source at a given frequency $\omega$ and position $\mathbf{r}'$~\cite{novotny}.

Starting from the dyadic Green's function in Eq.~(\ref{greens}), we define the LDOS~\cite{novotny}. This quantity represents the density of EM modes available in a given medium which a classical dipole can couple to~\cite{novotny}. The LDOS is defined as
%
%
%%%%%%%%%%%%%%
\begin{equation}\label{LDOS_mu}
    \rho_{\mu} (\mathbf{r};\omega)=\frac{6\omega}{\pi c^2} \left\{  \mathbf{n}_\mu \cdot \mathrm{Im} \left [ {\mathbf{G}}(\mathbf{r},\mathbf{r};\omega) \right] \cdot \mathbf{n}_\mu\right\}.
\end{equation}
%%%%%%%%%%%%%%
%
%
 with ${\bmu}=|{\bmu}|\mathbf{n}_\mu$ the dipole moment, where $\mathbf{n}_\mu$ denotes the unit vector in the $\mu$-direction. 
For the case of TM polarization, we obtain a finite LDOS when considering a dipole along the $z$ direction. For this case we can express the LDOS as:
%
%
%%%%%%%%%%%%%%
\begin{equation}\label{LDOS}
    \rho(\mathbf{r};\omega)=
    \frac{6\omega}{\pi c^2} \mathrm{Im} \left [ G_{zz} (\mathbf{r},\mathbf{r};\omega) \right]\, ,
\end{equation}
%%%%%%%%%%%%%%
%
%
where $G_{zz} (\mathbf{r},\mathbf{r};\omega)$ is the $zz$-component of the dyadic Green's function in Eq.~(\ref{greens}). 
We can simplify Eq.~(\ref{greens}) using the Sokhotski-Plemelj formula~\cite{stone_2009}. As a consequence of this simplification, the LDOS in Eq.~(\ref{LDOS}) will read
%
%
%%%%%%%%%%%%%
\begin{equation}\label{LDOS_cont}
    \rho (\mathbf{r};\omega)=\frac{6}{\pi S}
     \int_\mathrm{BZ} d\mathbf{k}\, \varepsilon(\mathbf{r})\sum_{n}
     |\mathbf{E}_{\mathbf{k},n}(\mathbf{r})|^2
%    \mathbf{E}_{\mathbf{k}n}(\mathbf{r}) \otimes \mathbf{E}_{\mathbf{k}n}^*(\mathbf{r}) 
\delta(\omega - \omega_{\mathbf{k}n})\, .
\end{equation}
%%%%%%%%%%%
%
%

We obtain the DOS by integrating Eq.~(\ref{LDOS}) over the UC of the system
%
%
%%%%%%%%%%%%%%%
\begin{eqnarray}\label{DOS}
    J(\omega)& =\frac{6}{\pi S}
     \int_\mathrm{UC} d\mathbf{r}\,\int_\mathrm{BZ} d\mathbf{k}\, \varepsilon(\mathbf{r}) \sum_{n}
    |\mathbf{E}_{\mathbf{k}n}(\mathbf{r})|^2 \delta(\omega - \omega_{\mathbf{k}n}) \nonumber \\
    & = \frac{6}{\pi}\int_\mathrm{BZ} d\mathbf{k} \sum_{n}
    \delta(\omega - \omega_{\mathbf{k}n})\, .
\end{eqnarray}
%%%%%%%%%%%%%%%%%
%
%
Here, we assume that the quasi-transversal modes are normalized in the UC as~\cite{McPhedran_2004,Sakoda_2005}
%
%
%%%%%%%%%%%%%%%%
\begin{equation*}
\frac{1}{S}\int_\mathrm{UC}\varepsilon(\mathbf{r})|\mathbf{E}_{\mathbf{k}n}(\mathbf{r})|^2 d\mathbf{r}=1.
\end{equation*}
%%%%%%%%%%%%%%%
%
%

On the other hand, we can obtain the EM energy density by integrating  Eq.~(\ref{LDOS_cont}) in energy over the set bands $\Lambda$ of interest
%
%
%%%%%%%%%%%%%%%
\begin{eqnarray}\label{density}
    n_\Lambda(\mathbf{r})& =\frac{1}{S}
     \int_\Lambda d\mathbf{\omega}\, \int_\mathrm{BZ} d\mathrm{k}\,\varepsilon(\mathbf{r}) \sum_{n\in\Lambda}
    |\mathbf{E}_{\mathbf{k}n}(\mathbf{r})|^2 \delta(\omega - \omega_{\mathbf{k}n}) \nonumber \\
    & = \frac{6}{\pi S} \sum_{n\in\Lambda} \int_\mathrm{BZ} d\mathbf{k}\,
     \varepsilon(\mathbf{r}) 
    |\mathbf{E}_{\mathbf{k}n}(\mathbf{r})|^2 \,,
\end{eqnarray}
%%%%%%%%%%%%%%%%%
%
%
We note in passing that, starting from this last expression, we can obtain the total EM energy density by summing it over all disjoint sets of bands $\Lambda$, \emph{i.e.} $n(\mathbf{r})=\sum_\Lambda n_\Lambda(\mathbf{r})$. The last quantity can be expressed in terms of EM field Wannier functions \cite{busch2003wannier,albert2000generalized,albert2002photonic,busch2011photonic,gupta2022wannier} $E^w_{n\mathbf{R}}(\mathbf{r})$~(\ref{BlochToWannier}) so that the total EM energy density reads:
%
%
%%%%%%%%%%%%%%
\begin{equation}\label{ted}
    n(\mathbf{r})=\frac{6}{\pi S}\sum_\Lambda \sum_{n\in\Lambda} \sum_{\mathbf{R}} \varepsilon(\mathbf{r}) |E^w_{n\mathbf{R}}(\mathbf{r})|^2\,.
\end{equation}
%%%%%%%%%%%%%%
%
%
This expression gives us an indirect access to the Wannier modes of the system without the need to evaluate them directly~\cite{wolff2013generation}.

In the following, we describe how the LDOS, DOS, and the EM energy density expressions defined in Eqs.~(\ref{LDOS}-\ref{density}) can be numerically evaluated starting from the Bloch states of the PhC. 
As a first step, noting that we consider only 2D PhCs, we can transform the surface integrals containing a delta function in Eqs.~(\ref{LDOS}-\ref{density}) into line integrals as follows~\cite{Sakoda_2005,ashcroft_1976}:
%
%
%%%%%%%%%%%%%%
\begin{eqnarray}\label{OSlines}
    \rho(\mathbf{r};\omega)& = \frac{6}{\pi S}\sum_n
     \int_{\ell_{\mathbf{k}n}} d\ell\, \varepsilon(\mathbf{r}) 
    \frac {\left | \mathbf{E}_{\mathbf{k}n}(\mathbf{r}) \right |^2} {\left |\mathbf{\nabla_k} \omega_{\mathbf{k}n} \right|}, 
    \label{LDOS_line}    \\
    %%%%%%%%%%%%%%%%%%%%%%
    %%%%%%%%%%%%%%%%%%%%%%
    %%%%%%%%%%%%%%%%%%%%%%
    J(\omega)&=\frac{6}{\pi}\sum_n\int_{\ell_{\mathbf{k}n}} 
    d\ell\, \, \varepsilon(\mathbf{r})\frac 1 {\left |\mathbf{\nabla_k} \omega_{\mathbf{k}n} \right|} \, ,
    %\!\!=\!\! \sum_n\int_{\ell_{\mathbf{k},n}}\!\! d\ell 
    %\frac 1 {\left | \mathbf{{v}}(\mathbf{k})  \right|},
    \label{DOS_line} \\
    %%%%%%%%%%%%%%%%%%%%%%
    %%%%%%%%%%%%%%%%%%%%%%
    %%%%%%%%%%%%%%%%%%%%%%
    n_\Lambda(\mathbf{r})&=
    \frac{6}{\pi S}\int_\Lambda d\omega \sum_{n\in\Lambda}
     \int_{\ell_{\mathbf{k}n}} d\ell\, \varepsilon(\mathbf{r}) 
    \frac {\left | \mathbf{E}_{\mathbf{k}n}(\mathbf{r}) \right |^2} {\left |\mathbf{\nabla_k} \omega_{\mathbf{k}n} \right|}
    \label{density_line} 
\end{eqnarray}
%%%%%%%%%%%%%%
%
%
where $\nabla_{\mathbf{k}}\omega_{\mathbf{k}n}$ is the gradient of the frequency function in the band $n$ along the wavevector $\mathbf{k}$, which coincides with the group velocity~\cite{Sakoda_2005}, and 
$\ell_{\mathbf{k}n}$ represents the closed contour defined by the momenta satisfying the condition $\omega=\omega_{\mathbf{k}n}$. 

We can straightforwardly discretize the expressions in Eqs.~(\ref{LDOS_line}-\ref{density_line}) as follows
%
%
%%%%%%%%%%%%%%%%
\begin{eqnarray}
    \rho(\mathbf{r};\omega)&=
     \sum_n \sum_{i} \varepsilon(\mathbf{r}) 
    \frac {\left | \mathbf{E}_{\mathbf{k}_i n}(\mathbf{r}) \right |^2} {\left |\mathbf{\nabla_k} \omega_{\mathbf{k}_in} \right|}\Delta \ell_n^i,\label{LDOS_line_discr}\\
    %%%%%%%%%%%%%%%%%%%%%%%%%%%%%%
    %%%%%%%%%%%%%%%%%%%%%%%%%%%%%%
    %%%%%%%%%%%%%%%%%%%%%%%%%%%%%%
    J(\omega)
    &=\sum_n\sum_{i} \varepsilon(\mathbf{r})
    \frac {1} {\left |\mathbf{\nabla_k} \omega_{\mathbf{k}_i n} \right|}\Delta \ell_n^i, \label{DOS_line_discr} \\
    %%%%%%%%%%%%%%%%%%%%%%%%%%%%%%
    %%%%%%%%%%%%%%%%%%%%%%%%%%%%%%
    %%%%%%%%%%%%%%%%%%%%%%%%%%%%%%
    n_\Lambda(\mathbf{r})&=
    \int_\Lambda d\omega \sum_{n\in\Lambda} \sum_{i} \varepsilon(\mathbf{r}) 
    \frac {\left | \mathbf{E}_{\mathbf{k}_i n}(\mathbf{r}) \right |^2} {\left |\mathbf{\nabla_k} \omega_{\mathbf{k}_in} \right|}\Delta \ell_n^i,\label{density_line_discr}
\end{eqnarray}
%%%%%%%%%%%%%%%%%
%
%
where $\Delta \ell_n^i$ is the discretized line differential. This quantity will be different at each particular point along the closed path $\ell_{\mathbf{k}n}$ and it is defined as  
%
%
%%%%%%%%%%%%%%
\begin{equation}\label{diff}
  \hspace{-1cm}  \Delta \ell_n^i=\sqrt{(k_{n\alpha}^{i+1}-k_{n\alpha}^{i})^2+(k_{n\beta}^{i+1}-k_{n\beta}^{i})^2+2(k_{n\alpha}^{i+1}-k_{n\alpha}^{i})(k_{n\beta}^{i+1}-k_{n\beta}^{i})\cos\theta}\,,
\end{equation}
%%%%%%%%%%%%%
%
%
where $k_{n\alpha}$ and $k_{n\beta}$ are the energy dependent projections of the paths $\ell_{\mathbf{k}n}$ over the two possible linearly independent momenta $(\mathbf{k}_\alpha,\mathbf{k}_\beta)$ in the 2D reciprocal space and $\theta$ is the angle between them.

On the practical side, we proceed in the following way: we first fix a grid discretization of the first BZ.\footnote{We verified that with $128\times128$ points, we obtain already converging results. This choice depends on the system; more points are needed to achieve numerical convergence for highly complex systems.}
We determine the energy isosurface $\ell_{\mathbf{k}n}$ in the discretized first BZ and compute the differential in Eq.~(\ref{diff}).\footnote{If the grid is not fine enough we can make an interpolation between the $\mathbf{k}$ points of the contour.} Finally, at each point in $\ell_{\mathbf{k}n}$ we compute the electric field and the spectrum gradient to apply Eqs.~(\ref{LDOS_line_discr}) and~(\ref{DOS_line_discr}).

Although these local observables were first derived to couple the PhC to an external dipole~\cite{McPhedran_2004}, in this article we will explore to what extent the EM energy density can be used to probe topological properties of bulk PhCs. In various examples of PhCs we show later, we will explore where the EM energy density concentrates from a set of bands in real space. Additionally, we also relate the position of the maximum EM energy density within the UC with the one predicted by the MLWFs. 

\section{The Breathing Honeycomb Lattice}\label{ex_One}

\subsection{PhC Model}

In this Section we characterize the topological properties of the breathing honeycomb PhC introduced by Wu and Hu in Ref.~\cite{wuandhu}. 
This is  defined as a two-dimensional (2D) honeycomb PhC with an artificially enlarged unit cell in the $(x,y)$-plane | where $z$ is the invariant direction. 
%
%
%%%%%%%%%%%%%%%    
\begin{figure*}[!ht]
\begin{flushright}
\includegraphics[width=0.85\textwidth]{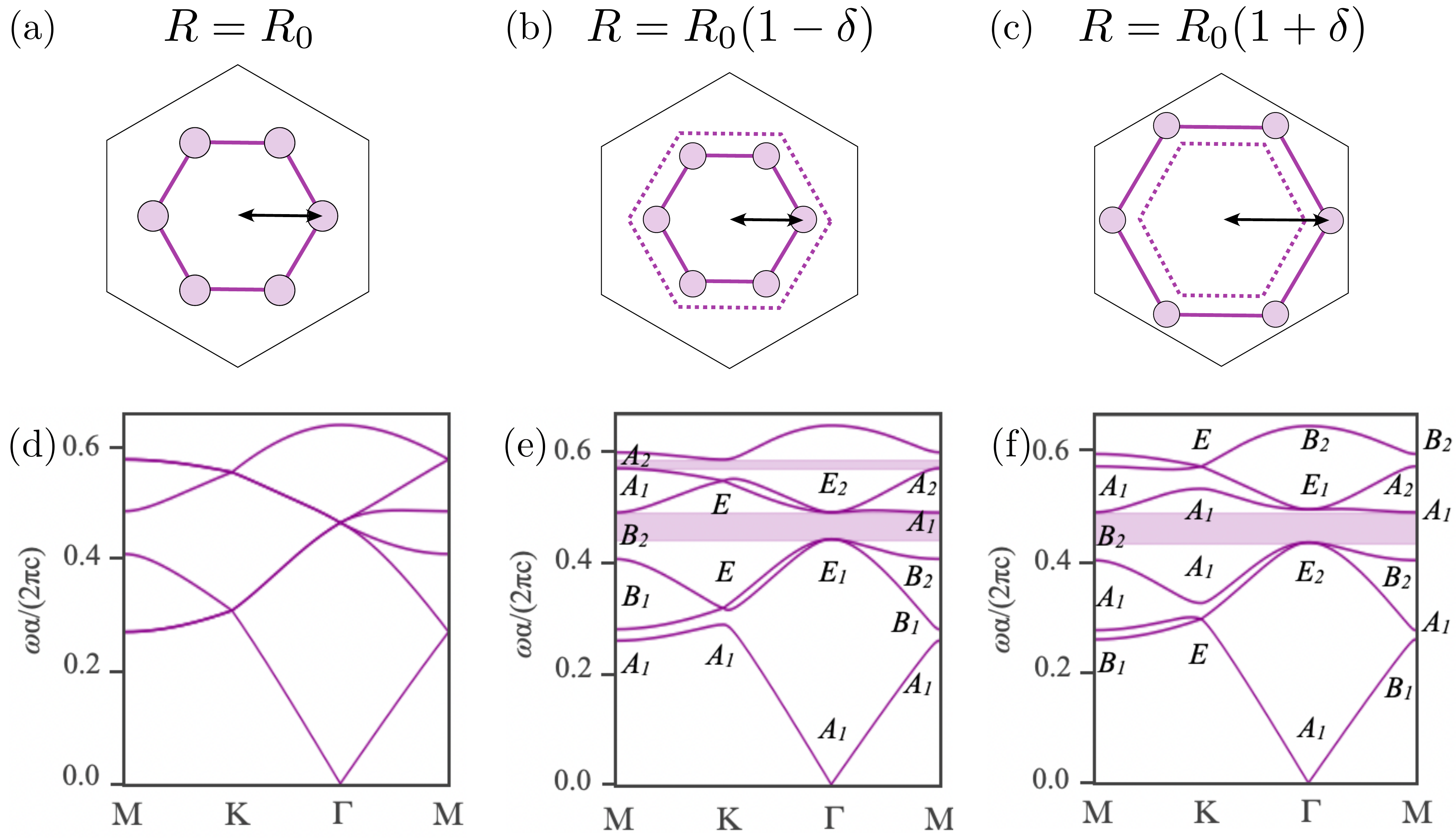}
\end{flushright}
\caption{Unit cell of the breathing honeycomb lattice. Each cell is composed of six silicon rods ($\varepsilon = 11.7$) of radius $r = 0.12a$ in vacuum ($\varepsilon = 1$). (a) case of the unperturbed honeycomb lattice with the rods placed at a distance $R_0 = a/3$ from the origin. (b) case of the contracted lattice whose cylinders are moved towards the center at a distance $R = R_0(1 - \delta)$, with the perturbation $\delta= 0.11$. (c) case of the expanded honeycomb lattice with the rods displaced to a distance $R = R_0(1 + \delta)$ from the center, keeping the perturbation $\delta= 0.11$.
The TM modes frequency bands are shown in Panels (d) for $\delta=0$, (e) for $\delta=-0.11$, and (f) for $\delta=+0.11$. For contracted and expanded configurations a gap opens and it is indicated by a shaded area. For these two cases we indicate the \emph{irrep} of each Bloch modes at high-symmetry $\mathbf{k}$-points, as well.
}\label{figure_1}
\end{figure*}
%%%%%%%%%%%%%%%
%
%
Its non-primitive unit cell is composed of six high-purity silicon rods ($\varepsilon = 11.7$)~\cite{wuandhu,Dunlap_1953} of radius $r = 0.12a$ in vacuum ($\varepsilon = 1$), $a$ being the lattice constant. Each of these rods is located at a distance $R = R_0(1 \pm \delta)$ from the origin, where $R_0 = a/3$ is the location of the cylinders in the unperturbed honeycomb arrangement. The parameter $\pm \delta$ quantifies the breathing in the system which expands or shrinks the the hexagon formed by the six cylinders in the unit cell. Assuming $\delta>0$, the lattice is expanded for $+\delta$, whereas it is contracted for $-\delta$, as is shown in Figs.~\ref{figure_1}(a) to~\ref{figure_1}(c). 
In the absence of dispersion the problem scales with $a$. A value of the permittivity similar to $\varepsilon\approx12$ and negligible loss applies to Si for frequencies in a wide range, including the telecom band, for which typical values of the lattice constant are $a\sim \pi c/\omega\sim\mu$m  \cite{parappurath2020direct}.

The frequency band dispersion of the TM modes in the unperturbed honeycomb lattice presents an artificial four-fold degeneracy at $\mathbf{\Gamma}$. This degeneracy is associated to the  band structure folding due to the non-primitive unit cell as shown in Fig.~\ref{figure_1}(d). We consider a breathing of $\delta= \pm 0.11$, that moves the rods farther from ($\delta= + 0.11$) or closer to ($\delta= - 0.11$) the center of the unit cell. The TM band structure of both perturbed lattices shows the opening of a band gap between $\omega a/(2\pi c) = 0.4 - 0.5$; the two cases are related by a change in the character of the gap, indicating the presence of a topological phase transition. The band structures for the contracted and expanded PhC are shown in  Figs.~\ref{figure_1}(e) and~\ref{figure_1}(f), respectively. 

\subsection{Topological analysis}
%
%
%%%%%%%%%%%%%%
\begin{table}[!b]
\centering
\begin{tabular}{c||c|c|c}
EBR & $\Gamma$ & $K$ & $M$\\
\hline
\hline
$(A_1\uparrow G)_{1a}$ & $A_1$ & $A_1$ & $A_1$ \\
$(E_1\uparrow G)_{1a}$ & $E_1$ & $E$ & $B_1\oplus B_2$ \\
\hline\hline
$(A_1\uparrow G)_{3c}$ & $A_1\oplus E_2$ & $E\oplus A_1$ & $A_1\oplus B_1\oplus B_2$
\end{tabular}
\caption{\label{table_1}Relevant EBRs of $p6mm$ for the lowest three bands of the photonic breathing honeycomb lattice.}
\end{table}
%%%%%%%%%%%%%%
%
%
We start by determining the topological properties of the system by applying TQC. In Figs.~\ref{figure_1}(e) and~\ref{figure_1}(f) we present the band structure of the contracted and expanded lattices, respectively, together with the \emph{irreps} of the Bloch modes at the high-symmetry $\mathbf{k}$-points, computed using GT-Pack~\cite{gtpack1,gtpack2}.
We use the catalogue of Elementary Band Representations (EBRs) of the space group of the lattice ($p6mm$)~\cite{Bilbao1,Bilbao2,Bilbao3}, to characterize the topology of these two gapped systems. We summarize the relevant EBRs for the breathing honeycomb lattice in Table~\ref{table_1}.

For topologically trivial gapped systems, the set of little group representations of connected bands can be expressed as a linear combination of these EBRs with positive coefficients. Therefore, we can identify where the MLWFs  that induces these connected bands must be located in the UC. 
For both gapped systems, the expanded and contracted lattice, we consider the three lowest frequency bands for the corresponding analysis, which is shown in Table~\ref{table_2}.
%
%
%%%%%%%%%%%%%%
\begin{table}[!t]
\centering
\begin{tabular}{c||c|c|c|c}
 & $\Gamma$ & $K$ & $M$ & EBR\\
 \hline \hline
  Expanded & $A_1,E_2$ & $E,A_1$ & $B_1,A_1,B_2$ & $(A_1\uparrow G)_{3c}$\\
  Contracted & $A_1,E_1$ & $A_1,E$ & $A_1,B_1,B_2$ & $(A_1\uparrow G)_{1a} \oplus (E_1\uparrow G)_{1a}$\\
\end{tabular}
\caption{\label{table_2}Little group \emph{irreps} for the three lowest bands of each phase. Together with the corresponding EBR for the contracted lattice, or sum of EBRs for the expanded lattice. }
\end{table}
%%%%%%%%%%%%%%
%
%
    
We observe at $\mathbf{\Gamma}$ that the \emph{irrep} of bands 2~\&~3 is $E_1$ for the contracted lattice and $E_2$ for the expanded one. These changes in \emph{irrep} labels indicate a modification of the Bloch modes' character in each structure for the lowest  set of frequency  bands. At the \textbf{K}-point, we observe that in the contracted case there is a degeneracy between bands 2~\&~3 and band $1$ is isolated, whereas, in the expanded lattice, bands $1~\&~2$ are degenerate while band $3$ is not. Therefore, the \emph{irreps} at the \textbf{K}-point flip their character from $A_1, E$ in the contracted case to $E, A_1$ for the expanded lattice. Similarly, at the \textbf{M}-point, the \emph{irreps} of the bands $1~\&~2$ are flipped, being $A_1, B_1$ for the contracted lattice and $B_1, A_1$ for the expanded case. 
For the contracted lattice we observe two sets of isolated bands, band $1$ and bands $2~\&~3$, while for the expanded lattice we only observe one connected set formed by the three lowest bands. We use this information to determine which EBRs form the band representation of each set of connected bands. From the EBRs, we will gain the knowledge on the \emph{irrep} of the MLWFs which induce the bands. Additionally, we determine their location in real space, this will be labelled by one of the WPs.

From the EBRs of the contracted lattice (Tab.~\ref{table_2}), we observe that the two sets of bands can be induced from MLWFs placed at the position $1a$. This position corresponds to the center of the photonic particle | Fig.~\ref{figure_1}(b). The lowest energy band is induced from a Wannier function of character $A_1$ and the second set is induced from Wannier functions of character $E_1$. Therefore, for the the contracted lattice, the Wannier functions sit at the position $1a$ | we call this limit the natural (atomic) limit. On the contrary, for the expanded lattice, the three lowest frequency bands form a connected set  induced from a set of Wannier functions centered at $3c$ and each transforming under the \emph{irrep} $A_1$ of the site symmetry group. The $3c$ position is at the edge between two consecutive UCs. This condition corresponds to a photonic OAL; usually, an obstructed phase is defined with respect to a natural limit~\cite{NaturePaper}. It is important to note that both limits admit an exponentially localized Wannier representation. Thus, we conclude that this system presents a trivial topological character. This is in  contradiction to Refs.~\cite{Wu2016,Wei_21} claiming that the breathing honeycomb lattice is a photonic $\mathbb{Z}_2$ topological insulator. In this respect, we can interpret these two configurations as a 2D analogy of the one-dimensional Su-Schrieffer-Heeger chain~\cite{NaturePaper,ssh3} or the breathing kagome lattice~\cite{Proctor_2021,Herrera_2022}. These breathing systems do not display any robust topological features, because the two possible phases always correspond to different atomic limits.

\begin{figure}[ht!]
\begin{flushright}
\includegraphics[width=0.85\columnwidth]{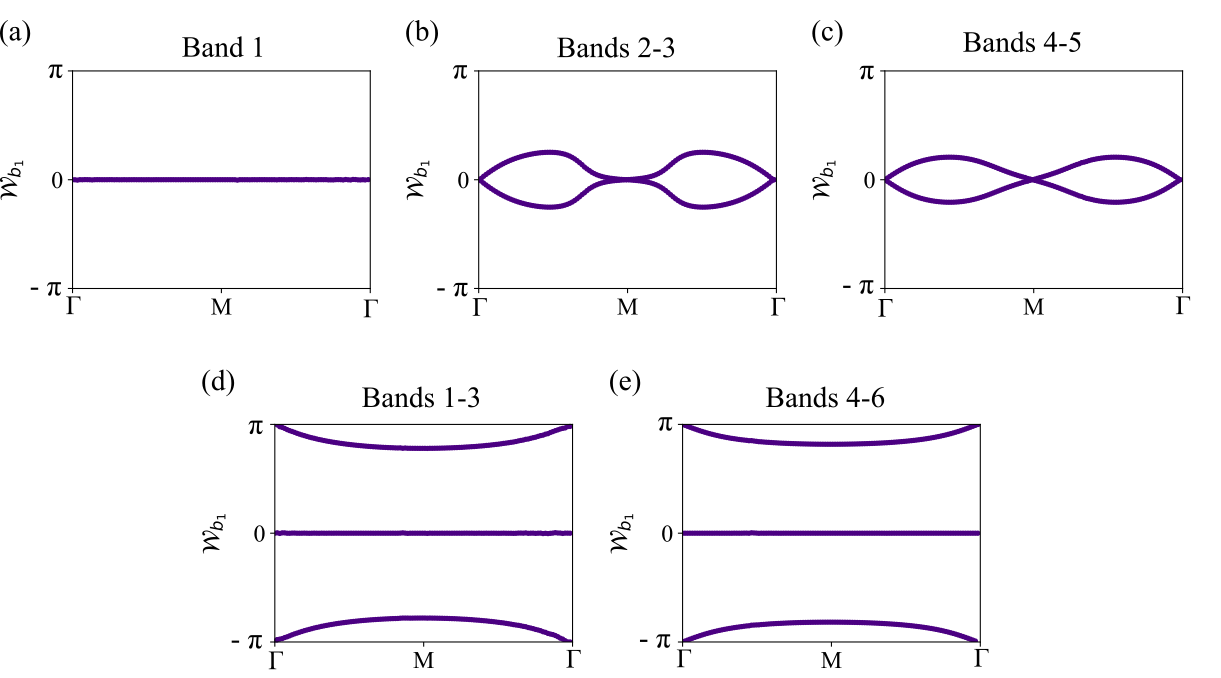}
\end{flushright}
\caption{Wilson-loop spectra along $\Gamma$--M--$\Gamma$ of each set of connected bands. For the case of the contracted lattice, in panel (a) to (c)  bands 1, 2-3 and 4-5, respectively. These spectra indicate that the Wannier centers are trivially localized around the center of the unit cell. For the expanded lattice, in panels (d) and (e) bands 1-3 and 4-6, respectively. Here we find that the spectra indicate that the Wannier centers as well localized at the edge of the unit cell, thus representing an obstruction phase of the trivial phase. Figure adapted from Supplemental Material of Ref.~\cite{Matt}.}\label{figure_2}
\end{figure}

In the following, we confirm the results obtained via TQC by calculating the WL spectrum for each set of connected bands. We present the results in  Fig.~\ref{figure_2},
%
%
%%%%%%%%%%%%%%%    
%%%%%%%%%%%%%%%
%
where we can clearly see that there is no winding in the WL spectra of the two phases of the breathing honeycomb lattice. Nevertheless, we  observe that the localization of the Wannier centers is different for each phase. 
For the contracted phase ($\delta=-0.11$), we confirm that the MLWFs are placed at the WP $1a$, which is reflected in the WL eigenvalues pinned near 0 | Figs.~\ref{figure_2}(a) to~\ref{figure_2}(c).
For the expanded case ($\delta=+0.11$), the MLWFs eigenvalues are not exclusively localized in the origin of the unit cell as for the contracted one, but they are localized around its edges as well. We can observe that at $\mathbf{\Gamma}$ the WL eigenvalues are pinned at $\pm \pi$ which indicates the MLWFs are shared between consecutive unit cells, \emph{i.e.} they are located at the $3c$ Wyckoff position, confirming that the system presents a photonic OAL phase.
%
%
%%%%%%%%%%%%%%%    
\begin{figure}[!ht]
\begin{flushright}
\includegraphics[width=0.85\columnwidth]{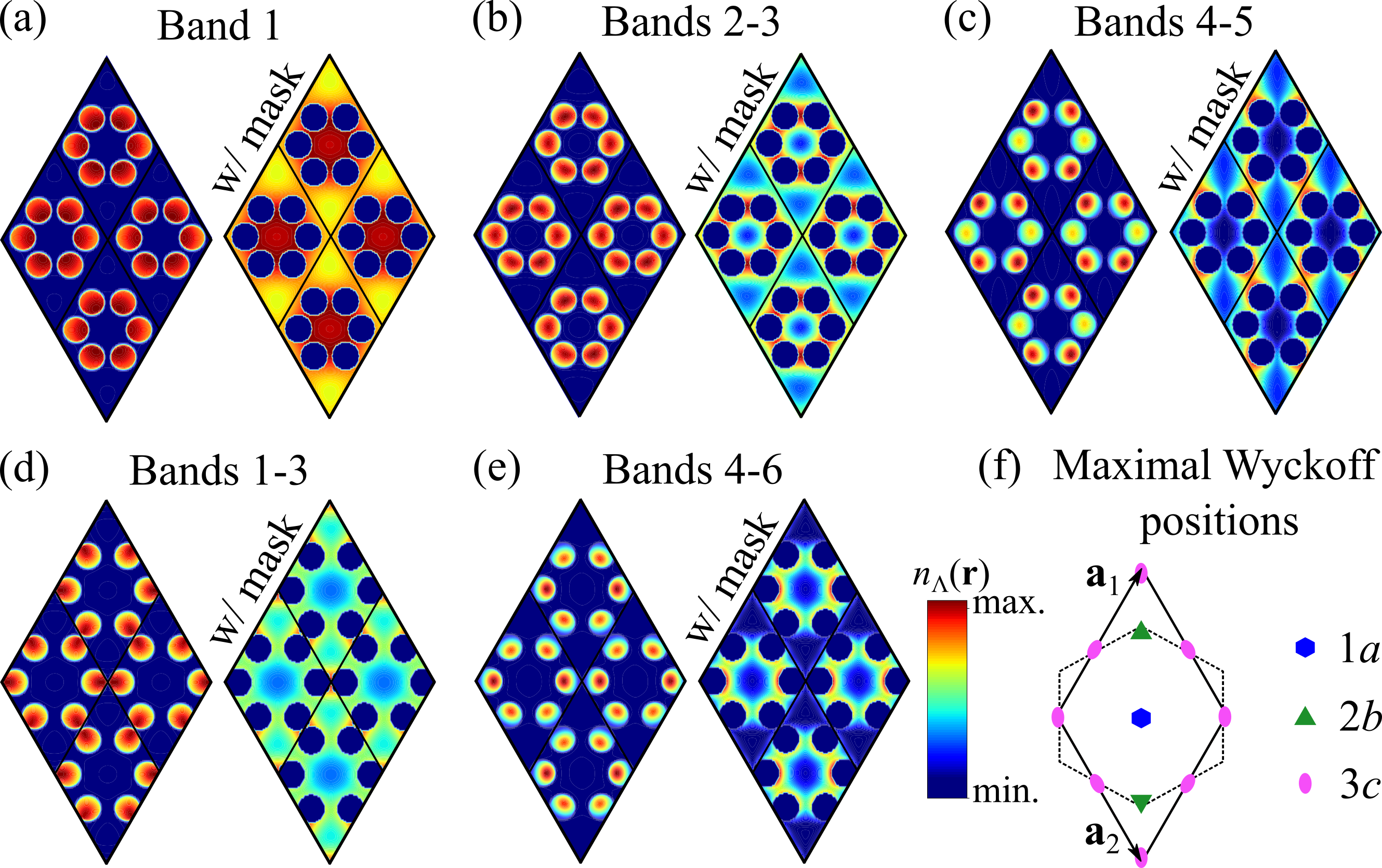}
\end{flushright}
\caption{\label{figure_3} Panels (a)-(e) show the EM energy density $n_\Lambda(\mathbf{r})$ associated to a set of connected bands. For each set of bands, $n_\Lambda(\mathbf{r})$ is evaluated for $\varepsilon = 11.7$ as shown on the left side of each panel, and those with a mask covering the rods at the right side. This procedure permits to resolve the EM energy density in the surrounding medium, air in our case. The legend is shown on the right part of each line, it has to be noted that the max[$n(\mathbf{r})$] in the medium is one order of magnitude larger than in air.  All the energy densities are plotted in a supercell composed of four consecutive unit cells. Panel (f) shows the unit cell at the top of the super cell  including the maximal Wyckoff positions as guide to the eye.} 
\end{figure}
%%%%%%%%%%%%%%%
%
%

To conclude our analysis, we use the EM energy density for characterizing each set of connected bands following the methodology described in Sec.~\ref{SecLocal}. We will show that the maximum values of the $n_\Lambda(\mathbf{r})$  can be associated  with the maximal Wyckoff position predicted with the previous two methods.

For the contracted lattice,  we observe that the EM energy density in the first band is mostly concentrated inside the dielectric rods and pointing towards the position $1a$ | Fig.~\ref{figure_3}(a). To better display the EM energy density in air, we artificially mask the $n_\Lambda(\mathbf{r})$ setting to zero its values inside the dielectric rods. It has to be noted that the EM energy density in air is 10-20 times smaller than inside the dielectric rods. The maximum of the $n_\Lambda(\mathbf{r})$ in the air region forms a ring centered around the origin and connecting the dielectric rods. We observe a highly symmetric EM energy density distribution which fits with the \emph{irrep} ($A_1$) of the band's EBR | see Table \ref{table_2}.

The EM energy density of the set of bands 2-3 shows a more precise localization of the maximum inside the dielectric rods with a less symmetric shape. The maximum values of the $n_\Lambda(\mathbf{r})$ presents an arc shape  connecting all the rods in a non-equal fashion | Fig.~\ref{figure_3}(b). Comparing it with the EM energy density of band $1$, we observe  a clear reduction of the symmetry, as indicated by the different EBR of the bands | which specifically is induced with $E_1$ \emph{irrep} in this case. Similar results are observed for the case of bands 4-5 where the EM energy density is still maximum inside the dielectric, the main difference is in the symmetry of the modes. This is especially visible for the case the of the masked EM energy density  | see Fig.~\ref{figure_3}(c).

For the expanded lattice, we observe that the EM energy density for the set of the three lowest frequency bands is localized inside the dielectric rods, but it points toward the position $3c$ | Fig.~\ref{figure_3}(d). Observing the $n_\Lambda(\mathbf{r})$ in air, its maximum values connect dielectric rods from different unit cells, \emph{i.e.}, the Wyckoff position $3c$. Moreover, we observe that the highly symmetric EM energy density is compatible with \emph{irrep} $A_1$, as indicated by the EBR for this set of connected bands | see Table \ref{table_2}. The next set of connected bands shows a deviation from the $3c$ localization similar to what observed for the set of bands 4-5 of the contracted configuration | Fig.~\ref{figure_3}(e). This lack of localization is indicative of the fact that the MLWFs for higher frequency bands have a longer localization length than those for the lower frequency bands.
%%%%%%%%%%

It is important to note that although both lattices present a trivial topology from the point of view of Wannier localization (TQC and WL analysis), the photonic OAL phase can emulate some non-trivial features. For example, since the $n_\Lambda(\mathbf{r})$ is maximally localized at the edges and corners of the unit cell, it allows for the emergence of symmetry-protected localized modes which can show robustness against certain types of disorder~\cite{Matt}.

\section{The Nested Breathing Honeycomb Lattice}\label{ex_Two}

This section explores the topological character of a new PhC that we construct by combining two copies of the breathing honeycomb lattice, one as expanded and one as contracted. The idea is to combine PhCs characterized by a  trivial and an OAL phase; the resulting PhC presents a rich phase diagram, including a fragile topological phase~\cite{Song_2020}. We name this lattice the nested breathing honeycomb lattice (NBHL).
%
%
%%%%%%%%%%%%%%%    
\begin{figure}[!t]
\begin{flushright}
\includegraphics[width=0.85\columnwidth]{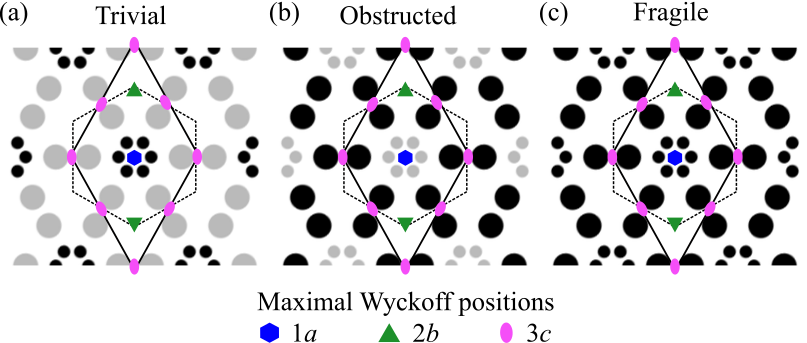}
\end{flushright}
\caption{\label{figure_4}The different photonic crystals configurations of the NBHL represented by three unit cells. Each unit cell is composed by six smaller dielectric rods of radius $r=0.05a$ placed at contracted honeycomb positions and other six rods of radius $r=0.1a$ placed at expanded honeycomb positions. The three configurations are characterized by different values of the dielectric constant $\varepsilon$ for the rods in the contracted and expanded positions | black rods with $\varepsilon = 12$ and grey rods $\varepsilon = 4$: trivial (a), OAL (b), and fragile (c). On the bottom, we mark the maximal Wyckoff positions of the space group $p6mm$. }
\end{figure}
%%%%%%%%%%%%%%%
%
%
\subsection{PhC Model}

The unit cell of the NBHL is composed by twelve dielectric rods, and is characterized by the space group $p6mm$.
Of the twelve rods, six are placed in the contracted honeycomb positions ($\delta= -0.11$), with radius $r=0.05a$, while the rest, with radius $r=0.1a$, take expanded honeycomb positions ($\delta= +0.11$). To explore the different configurations of the NBHL, we vary the dielectric constant $\varepsilon$ of the expanded part with respect to the contracted one. The different structures analyzed are shown in Figs.~\ref{figure_4}(a) to~\ref{figure_4}(c). In the following, we use rods with two distinct dielectric values:  $\varepsilon = 12$ (plotted in black) and  $\varepsilon = 4$ (plotted in grey). The first dielectric constant corresponds to $n$-type silicon~\cite{Kinasewitz_1983}, whereas the latter corresponds to BN~\cite{Lee_2018} at visible/near-IR frequencies.

We start by setting $\varepsilon = 12$ for the smaller rods located closer to the center of the unit cell, and $\varepsilon = 4$ for the larger rods, located closer to the edges  | we label this configuration ``Trivial" and we show it in Fig.~\ref{figure_4}(a). Since the contracted rods possess a higher $\varepsilon$, we expect that the EM energy density of low-lying frequency bands will be concentrated near the contracted rods.

We define the next structure setting $\varepsilon = 12$ for the bigger rods placed at the edges of the unit cell, and $\varepsilon = 4$ for the smaller rods closer the center. This configuration is labelled as ``Obstructed" in Fig.~\ref{figure_4}(b). The maximum localization of the fields is now expected around the rods in expanded positions, as they possess the highest~$\varepsilon$.

The last structure is defined by  $\varepsilon = 12$ for all the rods. We label this configuration as ``Fragile" and we show it in Fig.~\ref{figure_4}(c). 

In the following sections, we discuss the spectrum of the three different configurations together with our analysis of the topological character of each system using TQC, the eigenvalues of the WL operator and the EM energy density. 

\subsection{Trivial configuration}
\begin{figure}[!t]
\begin{flushright}
\includegraphics[width=0.85\columnwidth]{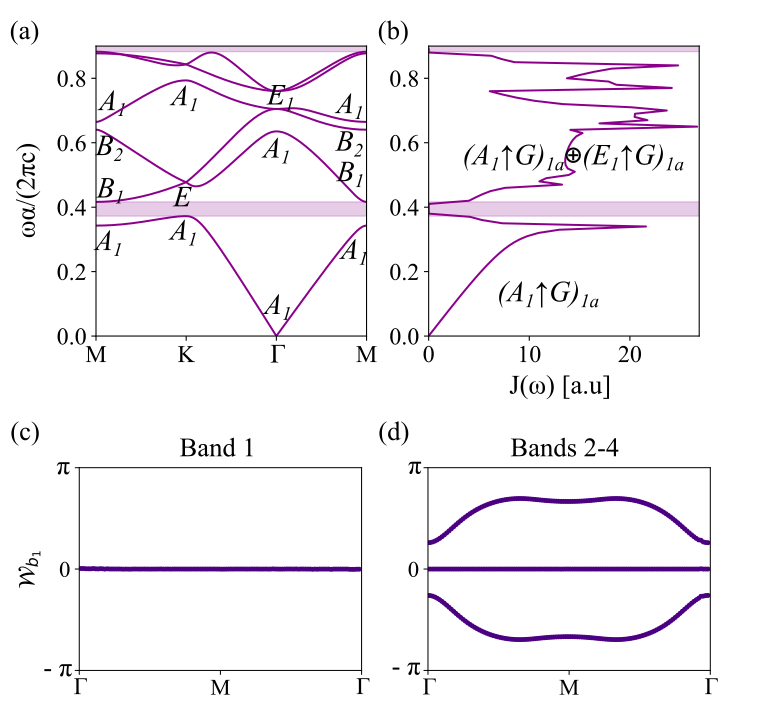}
\end{flushright}
\caption{\label{figure_5} Symmetry analysis and WL spectra of the trivial configuration. Panel (a) shows the band structure for TM modes with the corresponding \emph{irrep}  of each Bloch mode at high-symmetry $\bf{k}$-points labelled inset. The color code indicates different groups of connected bands. The correspond EBR of each set of bands is labelled inset. Panels (c) and (d) WL spectra for the band 1 and 2-4, respectively. }
\end{figure}

Here, we analyze the topological character of the first structure described above | Fig.~\ref{figure_4}(a).
In Figs.~\ref{figure_5}(a) and~\ref{figure_5}(b)  we show the spectrum with the \emph{irreps} and the DOS with the EBR of each set of connected bands labelled. In the DOS we can see the van~Hove singularities associated with the two-dimensional band structure. These singularities are related to $\mathbf{k}$-points where the group velocity of a band goes to zero | see Eq.~(\ref{DOS_line}). 
%
%
%%%%%%%%%%%%%%%    
%%%%%%%%%%%%%%%
%
First, we analyze the symmetry representation of the Bloch modes at high-symmetry $\bf{k}$-points. The collection of all the \emph{irreps} for each set of bands lets us identify a sum of EBRs consistent with the set of bands. For this configuration, we can express all the \emph{irreps} of the structure as positive linear combinations of EBRs of $p6mm$, meaning that the band structure is induced from exponentially localized MLWFs. The EBRs defining the first two sets of bands are collected in Table~\ref{table_3}.

The EBR of the first band, $(A_1\uparrow G)_{1a}$, indicates that this band is induced from a Wannier function maximally localized at the $1a$ Wyckoff position which transforms under the $A_1$ \emph{irrep}. The set of bands 2-4
has a MLWF localized at Wyckoff position $1a$ corresponding to the following EBR
$(A_1\uparrow G)_{1a}\oplus(E_1\uparrow G)_{1a}$.

%
%
%%%%%%%%%%%%%%%%%%
\begin{table}[!h]
\centering
\begin{tabular}{c||c|c|c|c}
 & $\Gamma$ & $K$ & $M$ & EBR\\
 \hline\hline
  Band 1 & $A_1$ & $A_1$ & $A_1$ & $(A_1\uparrow G)_{1a}$\\
  Bands 2-4 & $A_1,E_1$ & $E,A_1$ & $B_1,B_2,A_1$ & $(A_1\uparrow G)_{1a} \oplus (E_1\uparrow G)_{1a}$ \\
\end{tabular}
\caption{Little group \emph{irreps} for the set of lowest bands in the trivial structure; together with the corresponding EBR.}\label{table_3}
\end{table}
%%%%%%%%%%%%%%%%%%
%
%
As we observed from the EBRs, the functions which induced the band 1 and bands 2-3 are maximally localized at the $1a$ Wyckoff position. For the $p6mm$ space group, this position is located at the center of the unit cell. To support these results, we computed the eigenvalues of the WL operator | see Figs.~\ref{figure_5}(c) and~\ref{figure_5}(d). 

For the first band the WL eigenvalues are constant and equal to zero. As there is no winding, the Wannier function can be exponentially localized, thus  indicating the trivial character of the gap above. On the other hand, the value is related to the position of the MLWF. In this case, it corresponds to a MLWF at the center of the unit cell.

For the set of the bands 2-4, we observe no winding in the spectrum of the WL operator which indicates a trivial topological character, meaning that these bands are induced from exponentially localized Wannier functions. The WL spectrum can be continuously deformed to have all three bands at $0$, consistent with MLWFs at the $1a$ Wyckoff position.

\begin{figure}[!hb]
\begin{flushright}
\includegraphics[width=0.85\columnwidth]{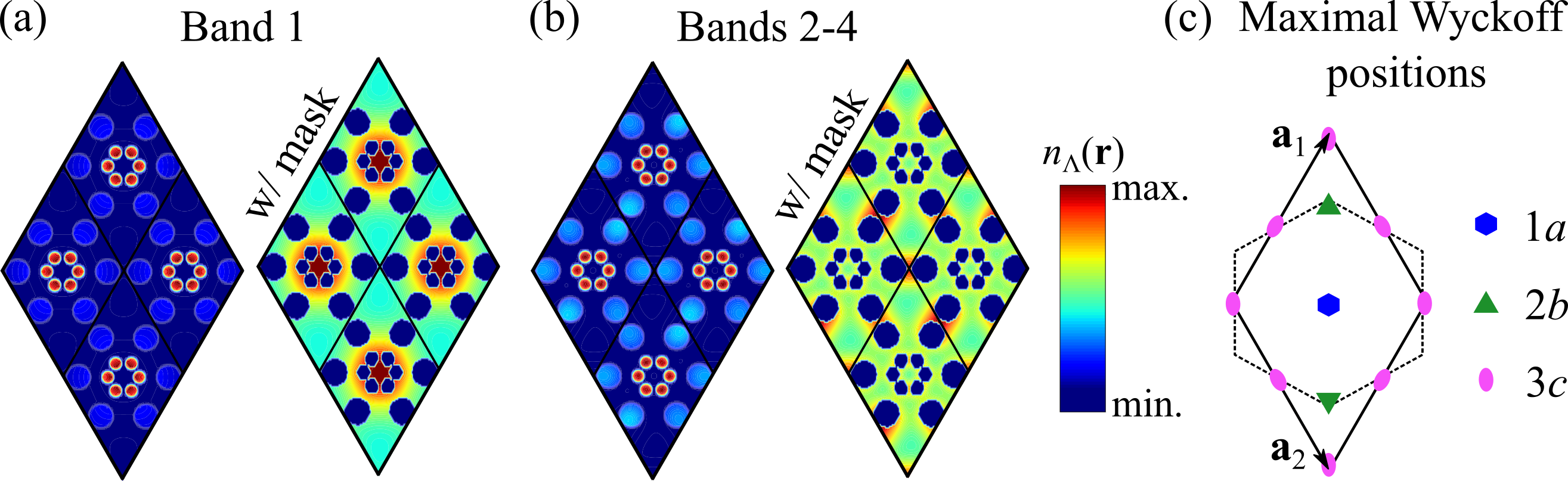}
\end{flushright}
\caption{\label{figure_6}EM energy density $n_\Lambda(\mathbf{r})$ associated to band 1 (a) and the set of connected bands 2-4 (b).
The $n_\Lambda(\mathbf{r})$ is calculated for ($\varepsilon_{\mathrm{cont.}}=  12, \varepsilon_{\mathrm{exp.}}= 4$) on  the  left  side of each panel, and those with a mask covering the rods at the right side. The legend is shown on the right part of each line, it has to be noted that the max[$n_\Lambda(\mathbf{r})$] in the medium is one order of magnitude larger than in air.  All the energy densities are plotted in a supercell composed of four consecutive unit cells. Panel (c) show the unit cell at the top of the super cell  including the maximal Wyckoff positions as guide to the eye.
}
\end{figure}

Finally, we complete this analysis by calculating the EM energy density for the different sets of bands | see Fig.~\ref{figure_6}. For clarity, the results plotted in a region containing four unit cells.
We perform the calculation considering $\varepsilon_{\mathrm{cont.}} = 12$ and $\varepsilon_{\mathrm{exp.}} = 4$, and  complement this result by displaying $n_\Lambda(\mathbf{r})$ in air on a separate scale.  

%
%
%%%%%%%%%%%%%%%    
%%%%%%%%%%%%%%%
%
%

For band 1,  we observe that the maximum of $n_\Lambda(\mathbf{r})$ is located inside the rods with higher $\varepsilon$ placed at the contracted positions | see Fig.~\ref{figure_6}(a). Furthermore, the maximum of the EM energy density is well localized around the center of the unit cell, \emph{i.e.} the $1a$ Wyckoff position. Thus, we find an agreement between the maximum of EM energy density with the Wannier center position predicted by the EBR of the band. 

For the set of bands 2-4,  we also observe the highest EM energy density  inside the rods in the contracted positions | Fig.~\ref{figure_6}(b). In this case, compared with the first band, we observe a higher concentration of $n_\Lambda(\mathbf{r})$ within the rods at expanded positions. Analyzing the EM energy density in air, we observe that the $n_\Lambda(\mathbf{r})$ is partially concentrated around the contracted rods but its  maximum is found between consecutive unit cells, at the edges and corners  corresponding to the $3c$ Wyckoff position. Although we observe the maximum $n_\Lambda(\mathbf{r})$ concentration around the center due to the higher $\varepsilon$, the EM energy density of the medium is not what is expected from the prediction of TQC, which indicates that the center of the Wannier function that induces this set of bands is placed at the $1a$ position. Instead we observe that the EM energy density has contributions that can be associated both to the  $1a$ and $3c$ WPs, even if the $3c$ contribution is smaller than the $1a$. This suggests that the Wannier functions for bands 2-4 have non-negligible support on the $3c$ Wyckoff position, and hence are not as well localized as the Wannier function for band 1.

%%%%%%%%%%%%%%%
%
\subsection{Obstructed configuration}

Here we analyze the opposite structure, \emph{i.e.} the one with the rods of higher $\varepsilon$ placed at the expanded honeycomb positions | see  Fig.~\ref{figure_4}(b).  In this case, we identify two sets of connected bands |  bands 1-3 and bands 4-6 | that we will characterize independently. 

\begin{figure}[ht!]
\begin{flushright}
\includegraphics[width=0.85\columnwidth]{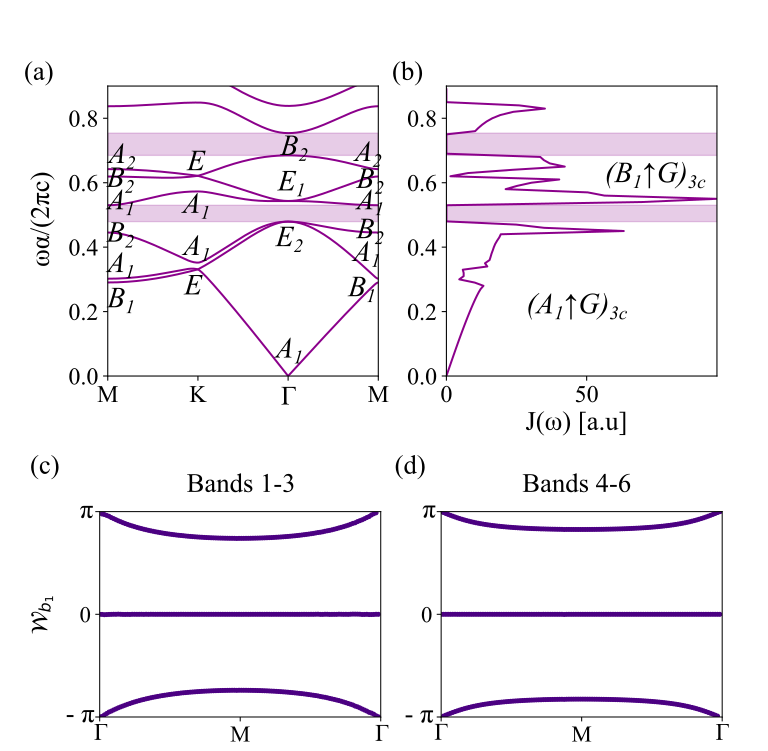}
\end{flushright}
\caption{\label{figure_7}Symmetry analysis and WL spectra of the obstructed configuration. Panel (a) shows the band structure for TM modes with the corresponding \emph{irrep} of each Bloch mode at high-symmetry $\bf{k}$-points labelled inset. The color shaded area indicates different groups of connected bands. The color code is kept for in panel (b), which shows the density of states in arbitrary units. The correspond EBR of each set of bands is labelled inset. Panels (c) and (c) spectra of the WL operator for the sets of bands 1-3 and bands 4-6.}
\end{figure}

In Fig.~\ref{figure_7}(a) we show the band structure with the \emph{irreps} at every high-symmetry points, and the corresponding DOS in Fig.~\ref{figure_7}(b). 
%
%
%%%%%%%%%%%%%%%    
%%%%%%%%%%%%%%%
%
%
The band representation of each set of bands and its relation with the EBRs of the $p6mm$ space group is summarized in Table~\ref{table_4}.
%
%
%%%%%%%%%%%%%
\begin{table}[ht!]
\centering
\begin{tabular}{c||c|c|c|c}
 & $\Gamma$ & $K$ & $M$ & EBR\\
 \hline\hline
  Bands 1-3 & $A_1,E_2$ & $E,A_1$ & $B_1,A_1,B_2$ & $(A_1\uparrow G)_{3c}$\\
  Bands 4-6 & $E_1,B_2$ & $A_1,E$ & $A_1,B_2,A_2$ & $(B_1\uparrow G)_{3c}$ \\
\end{tabular}
\caption{\label{table_4}Little group \emph{irreps} for the set of lowest bands in the obstructed limit structure; together with the corresponding EBR.}
\end{table}
%%%%%%%%%%%%%
%
%
The first set, composed by bands 1-3, is induced from a MLWFs centered at the $3c$ Wyckoff position that transforms under the \emph{irrep} $A_1$. The Wannier functions that induces the set of bands 4-6 are centered the $3c$ position as well, but transform under a different \emph{irrep} of the site symmetry group, labelled by $B_1$.

We now confirm these results by looking at the WL spectrum that we show in Figs.~\ref{figure_7}(c) and~\ref{figure_7}(d). 
For both set of bands we observe that the eigenvalues of the WL operator show no winding, indicating a trivial character. Nevertheless, the eigenvalues are located around $0$ and $\pm \pi$, indicating that the centers of the Wannier functions are located between consecutive unit cells, at the $3c$ position. Therefore, this structure is characterized as a photonic OAL phase. 
Although this phases can show features associated to topological effects such as corner and/or edge modes with some protection due to the lattice symmetries, the protection against disorder is not as robust as for a strong topological phases~\cite{Matt}.

\begin{figure}[ht!]
\begin{flushright}
\includegraphics[width=0.85\columnwidth]{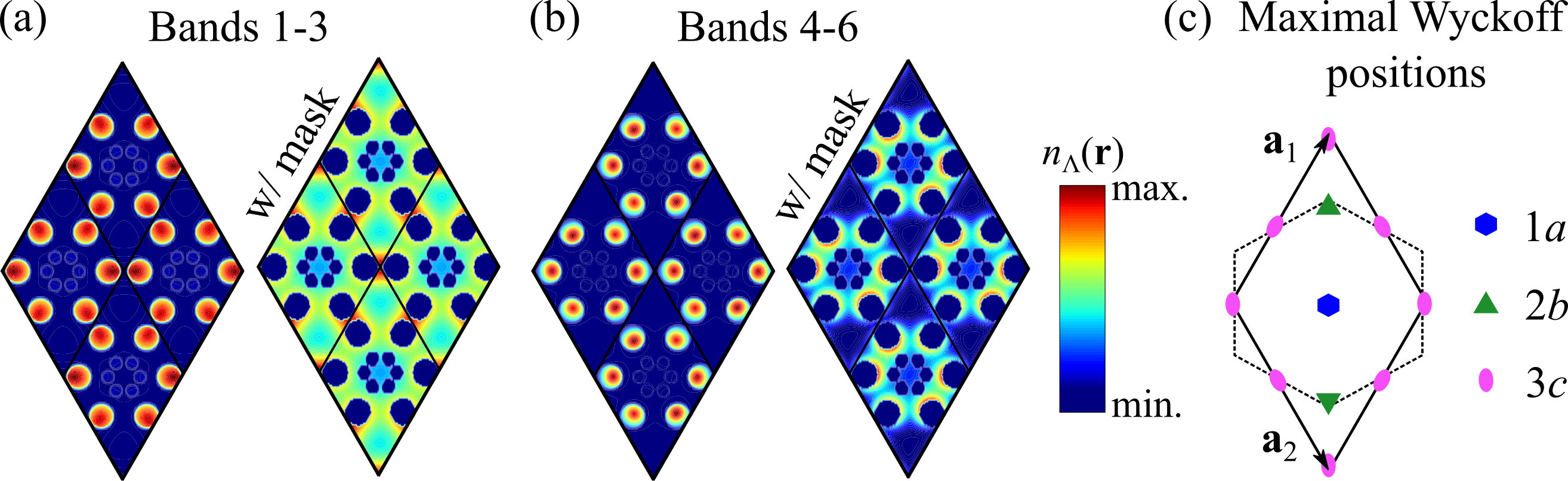}
\end{flushright}
\caption{\label{figure_8}
EM energy density $n_\Lambda(\mathbf{r})$ associated to set of band 1-3 (a) and the 2-4 (b).
The $n_\Lambda(\mathbf{r})$ is calculated for ($\varepsilon_{\mathrm{cont.}}=  4, \varepsilon_{\mathrm{exp.}}= 12$) on  the  left  side of each panel, and those with a mask covering the rods at the right side. The legend is shown on the right part of each line, it has to be noted that the max[$n_\Lambda(\mathbf{r})$] in the medium is one order of magnitude larger than in air.  All the energy densities are plotted in a supercell composed of four consecutive unit cells. Panel (c) show the unit cell at the top of the super cell  including the maximal Wyckoff positions as guide to the eye.}
\end{figure}

Finally, we characterize this PhC by exploring the EM energy density integrated over the sets of connected bands | see Fig.~\ref{figure_8}. 
%
%
%%%%%%%%%%%%%%%    
%%%%%%%%%%%%%%%
%
%
The $n_\Lambda(\mathbf{r})$ for the set of bands 1-3, shows a clear localization in the rods with higher dielectric constant with the maximum values oriented towards the edges of the unit cell. This result is confirmed by the EM energy density in air | see Fig.~\ref{figure_8}(a). 
Therefore, we find for these set of bands a good agreement between the predicted position of the Wannier centers ($3c$) and the position of the maximum values of $n_\Lambda(\mathbf{r})$. 
For the set of bands 4-6, we observe again the maximum within the rods with higher $\varepsilon$ placed at the expanded honeycomb positions. Compared to the previous set of bands, in this case the localization inside the rods is more defined and oriented to the center of the unit cell, confirmed by the $n_\Lambda(\mathbf{r})$ in air | see Fig.~\ref{figure_8}(b). This indicates that although the MLWFs for bands 4-6 are centered at the $3c$ position, they are less localized than the MLWFs for bands 1-3.

\subsection{Fragile configuration}

\begin{figure}[ht!]
\begin{flushright}
\includegraphics[width=0.85\columnwidth]{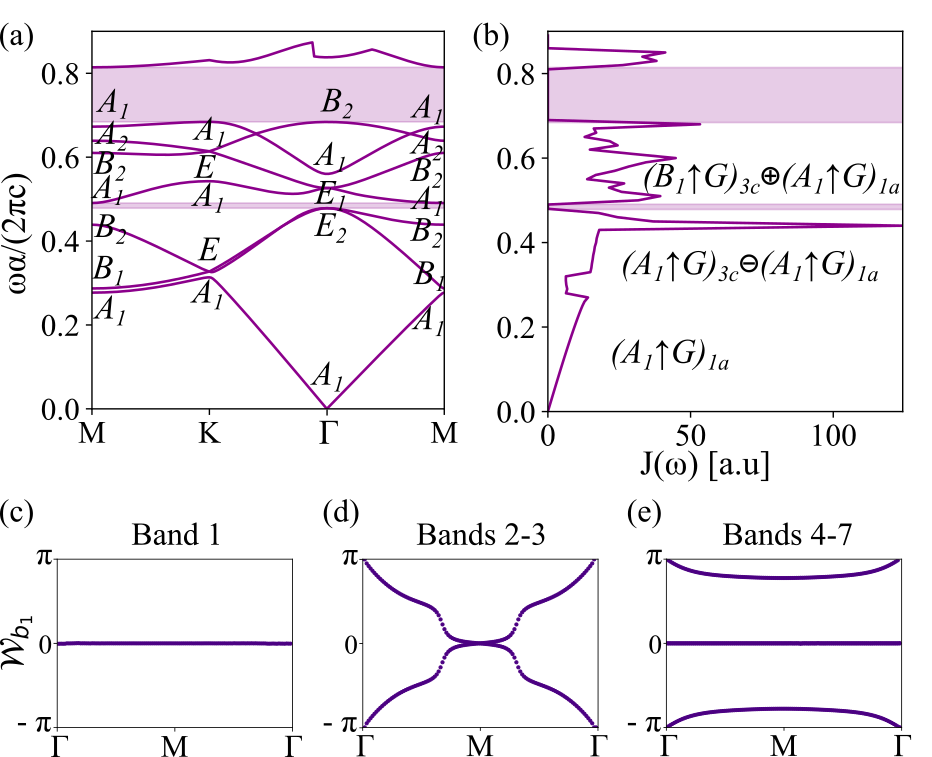}
\end{flushright}
\caption{\label{figure_9}Symmetry analysis and WL spectra of the fragile configuration. Panel (a) shows the band structure for TM modes with the corresponding \emph{irrep} of each Bloch mode at high-symmetry $\bf{k}$-points labelled inset. (b) the density of states  in arbitrary units. Each set of bands is labelled with the corresponding EBR. In this case, the band representation of the set of bands 2-3 | in increasing energy order | can be expressed as a subtraction of EBRs. (c) WL spectra of the sets of band 1 and (d) bands 4-7 shows no winding indicating trivial character, while the set of bands 2-3 (e) shows two windings with opposite slopes which indicates fragile topology.}
\end{figure}

To finish, we analyze the lattice composed of rods in expanded and contracted positions both with $\varepsilon = 12$ |   Fig.~\ref{figure_4}(c). For this structure, we distinguish three sets of bands: the first band isolated from the next set of bands 2-3 by a direct gap at each $\mathbf{k}$ (though there is no gap in the DOS); the set of bands 4-7 is completely isolated from all other bands in the DOS. We extract the \emph{irreps} at high symmetry points for each set of bands| labeled inset in Fig.~\ref{figure_9}(a), and compare them with the EBRs of space group $p6mm$. 
%
%
%%%%%%%%%%%%%%%    
%%%%%%%%%%%%%%%
%
%
We summarize the \emph{irreps} of each set of connected bands in Table \ref{table_5}. We include as well their corresponding EBRs of the space group $p6mm$. For this case, we observe that the set of bands 2-3 can only be expressed as a \emph{difference} of EBRs, indicating a fragile topological character. The sets of band 1 and bands 4-7 have trivial topological character since their band representation can be expressed as a single EBR and a sum of EBRs with positive coefficients, respectively. 

%
%
%%%%%%%%%%%%%%
\begin{table*}[!th]
\centering
\begin{tabular}{c||c|c|c|c}
 & $\Gamma$ & $K$ & $M$ & EBR\\
 \hline\hline
 Band 1 & $A_1$ & $A_1$ & $A_1$ & $(A_1\uparrow G)_{1a}$\\
  Bands 2-3 & $E_2$ & $E$ & $B_1,B_2$ & $(A_1\uparrow G)_{3c} \ominus (A_1\uparrow G)_{1a} $ \\
  Bands 4-7 & $E_1,A_1,B_2$ & $A_1,E,A_1$ & $A_1,B_2,A_2,A_1$ & $(B_1\uparrow G)_{3c} \oplus (A_1\uparrow G)_{1a}$ \\
\end{tabular}
\caption{\label{table_5}Little group \emph{irreps} of the lowest bands for the structure with fragile bands. The band representation is expressed in terms of EBRs of the space group $p6mm$ in the last column.}
\end{table*}
%%%%%%%%%%%%%%
%
%

In Fig.~\ref{figure_9}(b), we show the DOS of this structure with the EBRs for each set of bands labelled. TQC predicts that the Wannier function which induces the first band is centered at the  $1a$ Wyckoff position; Bands 4-7, are induced from a function that transforms respectively as a  sum of EBRs centered at $3c$ and $1a$. Bands 2-3 share the symmetry labels with a difference of EBRs at the $3c$ and $1a$ position, and hence should not have a symmetric, localized Wannier description. Therefore, we explore the eigenvalues of the WL operator to get information about the position of the Wannier function within the unit cell | shown in Figs.~\ref{figure_9}(c),~\ref{figure_9}(d) and~\ref{figure_9}(e) . 

%
%
%%%%%%%%%%%%%%%    
\begin{figure}[!h]
\begin{flushright}
\includegraphics[width=0.85\columnwidth]{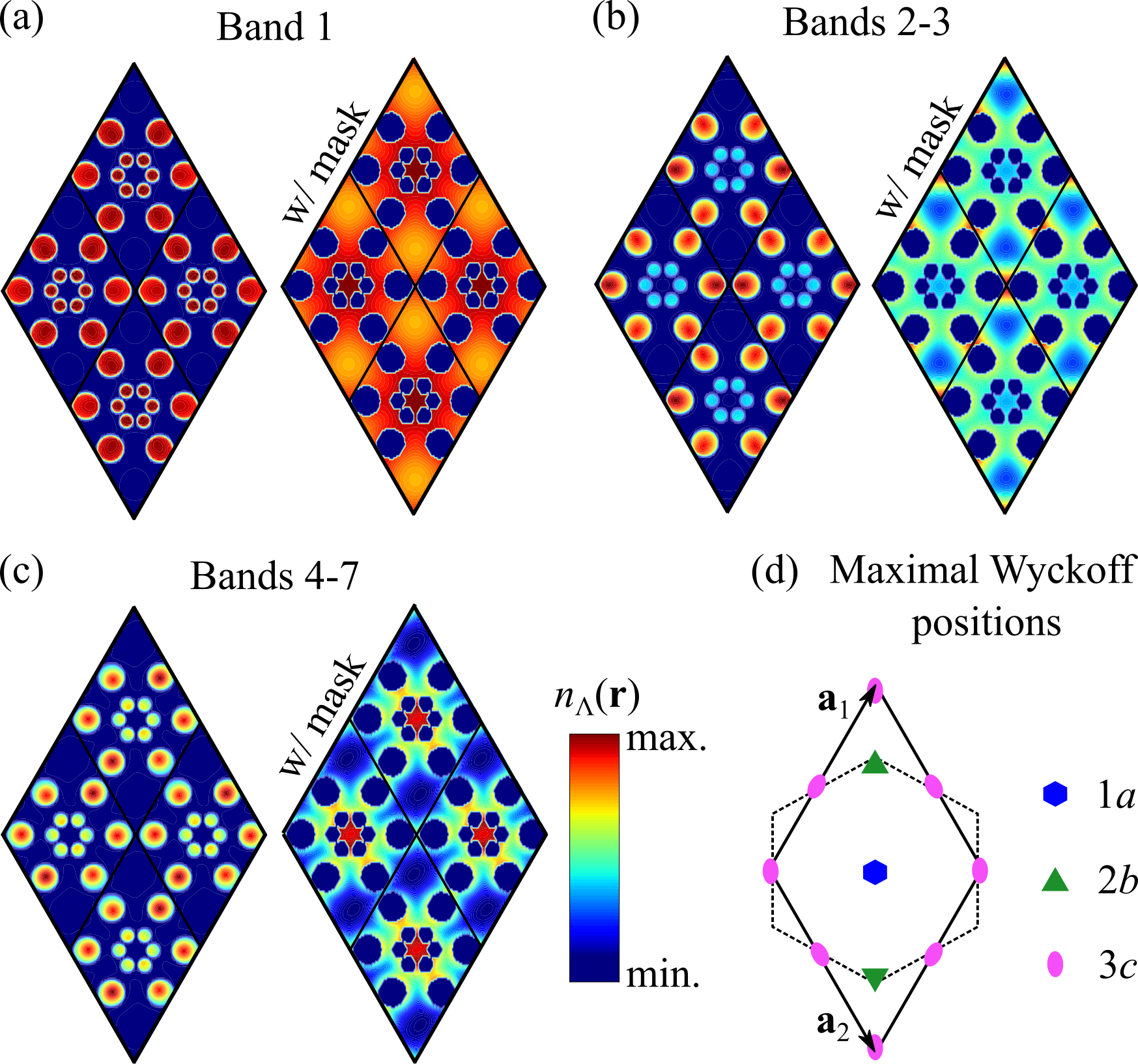}
\end{flushright}
\caption{\label{figure_10}
EM energy density $n_\Lambda(\mathbf{r})$ associated to band 1 (a), set of bands  2-4 (b) and 4-7 (c).
The $n_\Lambda(\mathbf{r})$ is calculated for ($\varepsilon_{\mathrm{cont.}}= \varepsilon_{\mathrm{exp.}}= 12$) on  the  left  side of each panel, and those with a mask covering the rods at the right side. The legend is shown on the right part of each line, it must be noted that the max[$n_\Lambda(\mathbf{r})$] in the air is one order of magnitude larger than in air.  All the energy densities are plotted in a supercell composed of four consecutive unit cells. Panel (d) show the unit cell at the top of the super cell  including the maximal Wyckoff positions as a guide to the eye.}
\end{figure}
%%%%%%%%%%%%%%%
%
%
The WL spectrum of the lowest energy band shows a constant zero value which indicates that the MLWF associated to this band is exponentially localized at the center of the unit cell, in the $1a$ Wyckoff position. 
For the set of bands 2-3, we get eigenvalues of the WL operator with a helical winding from $\pi$ to $-\pi$. This spectrum indicates that the Wannier functions that induce the bands are not exponentially localized at any position of the unit cell, signaling a non-trivial topology. As we explained in Sec.~\ref{WLSec}, this is the representative spectrum of $\mathbb{Z}_2$ insulators. Fragile topology can be distinguished from a $\mathbb{Z}_2$ topological phase by adding a trivial band to the calculation. While the $\mathbb{Z}_2$ phase preserves the winding, the fragile phase becomes trivial instead | with no winding in the WL spectrum.\footnote{Note that we have proven that the WL of the set of bands 1-3 is identical to the one of an OAL phase but for brevity we do not include this calculation in the manuscript~\cite{FragileLight}.}
The last spectrum, corresponding to the bands 4-7, does not present any winding and has bands centered around $\pm \pi$ and 0, which indicates that the Wannier functions are exponentially localized around the edge $3c$  and the center $1a$ positions of the unit cell. 

We conclude the analysis computing the EM energy density integrated over the different sets of bands, which allows us to explore the correlation with the predicted positions of Wannier functions that induce each set of bands. 

For the first band we observe the maximum $n_\Lambda(\mathbf{r})$ in the dielectric rods in contracted positions; while there is large EM energy density in 
the expanded rods as well and the EM energy density profile slopes toward the center of the unit cell (the $1a$ position)  | see Fig.~\ref{figure_10}(a) and  Fig.~\ref{figure_11} for cuts along different directions. 
The EM energy density for the set of bands 2-3 shows higher values within the rods in expanded positions oriented to the adjacent unit cells while the lowest values are within the rods in contracted positions. The same distribution is observed for the $n_\Lambda(\mathbf{r})$ in air | Fig.~\ref{figure_10}(b). 
For the last set of bands 4-7, the EM energy density shows the maximum values centered in the rods placed at expanded positions and lower values for those placed in the contracted positions. While $n_\Lambda(\mathbf{r})$ in air shows a clear maximum at the center of the unit cell and around the expanded rods pointing towards the center (the $1a$ position) Fig.~\ref{figure_10}(c).

\section{Discussion}\label{Outlook}
In the following, we discuss the implications of our results presented in Secs.~\ref{ex_One} and~\ref{ex_Two}.

We start from the breathing honeycomb lattice introduced in Ref.~\cite{wuandhu}. There has been a lot of theoretical and experimental activity studying the edge modes supported by this lattice~\cite{Wu2016,Siroki_2017,yves2017crystalline,gorlach2018far,Xia_2018,Wei_21,Barik_2018,Peng_2019,li2021experimental,Liu_2022}, and many works have claimed that this system is a strong $\mathbb{Z}_2$ photonic topological insulator. In Sec.~\ref{ex_One}, we have proven in a similar fashion to Ref.~\cite{Matt}, that the PhC based on the breathing honeycomb lattice is not a strong topological insulator but presents physics more similar to the Su-Schrieffer-Heeger model in one-dimension. In fact, it presents in-gap edge states characterized by a finite dispersion, but these edge states are completely contained within the gap without ever being degenerate with the photonic bands above and below~\cite{proctor2019exciting}. Thus, they are removable by perturbations. This physical aspect is intrinsic to the definition of the synthetic TRS operator introduced in Ref.~\cite{wuandhu}, that relies on a crystalline symmetry. As soon as this symmetry is broken, synthetic TRS is broken as well~\cite{Amo_2018}. However, as long as this symmetry is preserved, some of the topological features are still present and immune to perturbations as in a real strong topological system~\cite{Matt}.

The second aspect is the possible correlation between the position of the maximum of the EM energy density $n_\Lambda(\mathbf{r})$ and the position of the MLWFs deduced from TQC and the WL spectra. As can be observed in Figs.~\ref{figure_3},~\ref{figure_6}, and~\ref{figure_8}, we have obtained a good agreement between the position of the MLWFs as predicted by topological markers and the EM energy density. The correlation is more pronounced for low-frequency bands compared to the high-frequency ones, where we observed a systematic deviation of the direction towards which the EM energy density is pointing; this deviation is always stronger for the results in air compared to the exact EM energy density. 

However, we find that for the case of the PhC with the fragile configurations the agreement is very good.
%
%
%%%%%%%%%%%
\begin{figure}[!h]
    \begin{flushright}
    \includegraphics[width=0.85\columnwidth]{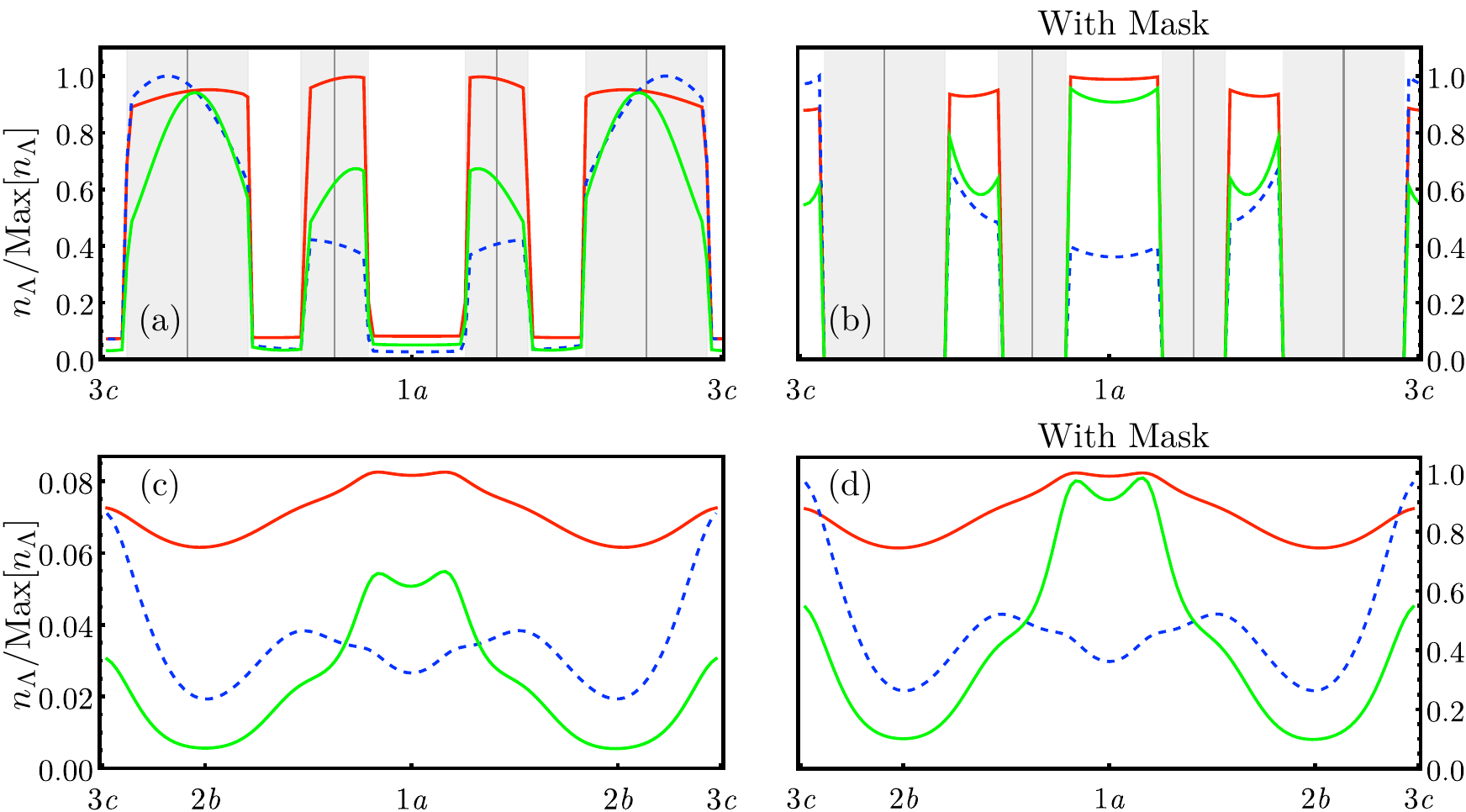}    
    \end{flushright}
    \caption{\label{figure_11}Cuts of the EM energy density in Fig.~\ref{figure_10} along $\mathbf{a}_1+\mathbf{a}_2$  and $\mathbf{a}_1-\mathbf{a}_2$ in Panels (a)-(b) and (c)-(d), respectively. Panels (b) and (d) refer to the case in which a mask covers the dielectric rods so to better display the contribution in air. In panels (a) and (b) the gray areas mark the presence of the dielectric rods with $\varepsilon(\mathbf{r})=12$ and the solid gray line is the center of dielectric rods. In all the panels $1a$, $2b$, and $3c$ indicate the maximal Wyckoff positions, the red-solid line corresponds to the band 1, the blue-dashed line to the band set 2-3, and the green-solid line to the band set 4-7.}
\end{figure}
%%%%%%%%%%%
%
%
In Fig.~\ref{figure_11} we show cuts of the EM energy density presented in Fig.~\ref{figure_10} along the two diagonals: one going through several dielectric rods and the WPs $1a$ and $3c$ in Fig.~\ref{figure_11}(a) and~\ref{figure_11}(b), and one going through the perpendicular direction that has no dielectric rods but passes through the WPs $1a$, $2b$, and $3c$.\footnote{We have investigated also cuts going parallel to the translation vectors and thus though different $3c$ Wyckoff positions and we have obtained similar conclusions.} In all panels we have normalized the EM energy density to its maximum value for the cases containing the dielectric rods in Fig.~\ref{figure_11}(a) and Fig.~\ref{figure_11}(c) or when masking the dielectric rods in Fig.~\ref{figure_11}(b) and Fig.~\ref{figure_11}(d).
For the case in Fig.~\ref{figure_11}(a) and~\ref{figure_11}(b), we observe that the largest value of the EM energy density for band 1 (red-solid line) is mostly localized inside the dielectric rods, but there is a general shift of all the maxima towards the $1a$ Wyckoff position\footnote{For the case in Fig.~\ref{figure_11}(c) the EM energy density is so small because no dielectric rods are contained along this path.}. We reach the same conclusions for the EM energy density along the perpendicular direction in Fig.~\ref{figure_11}(c) and~\ref{figure_11}(d), where we clearly see that the EM energy density is largest in the vicinity of Wyckoff position $1a$ as compared to the other two | its value is slightly increasing departing from the $1a$ position since close by points are passing near to two dielectric rods.%,  is largest in the vicinity of Wyckoff position $1a$ as compared to the other two.

The EM energy density for the set of  fragile bands (blue-dashed lines in Fig.~\ref{figure_11}) presents a maximum that is larger in the dielectric rods, pointing now to the $3c$ Wyckoff position. This result is in agreement with the fragile nature of this set of bands. The set can be trivialized by adding a trivial band induced by a $1a$ Wyckoff position. As a consequence, the new band set would be a photonic OAL induced by MLWFs placed in a $3c$ Wyckoff position.

Similar analysis can be done for the set of bands 4-7 (green-solid lines in Fig.~\ref{figure_11}) that present maxima well localized inside the dielectric but pointing towards both the $1a$ and the $3c$ Wyckoff position.

An alternative interpretation of the results in Fig.~\ref{figure_11}, can be obtained by considering that the EM energy density is proportional to the modulus square of the Wannier functions, see Eq.~(\ref{ted}). In this respect, the behaviour of the EM energy density of the set of bands 2-3 and 4-7 could be interpreted as the result of destructive  and constructive interference, respectively. 

From an experimental point of view, the EM energy density or other local observables such as the LDOS could be accessed by coupling the PhC to a quantum emitter \cite{perczel2020topological,navarro2021photon} or with electron probe spectroscopy~\cite{Peng_2019,Rivera_2020,Shuwai_2022}.

\begin{figure}[!t]
    \begin{flushright}
    \includegraphics[width=0.85\columnwidth]{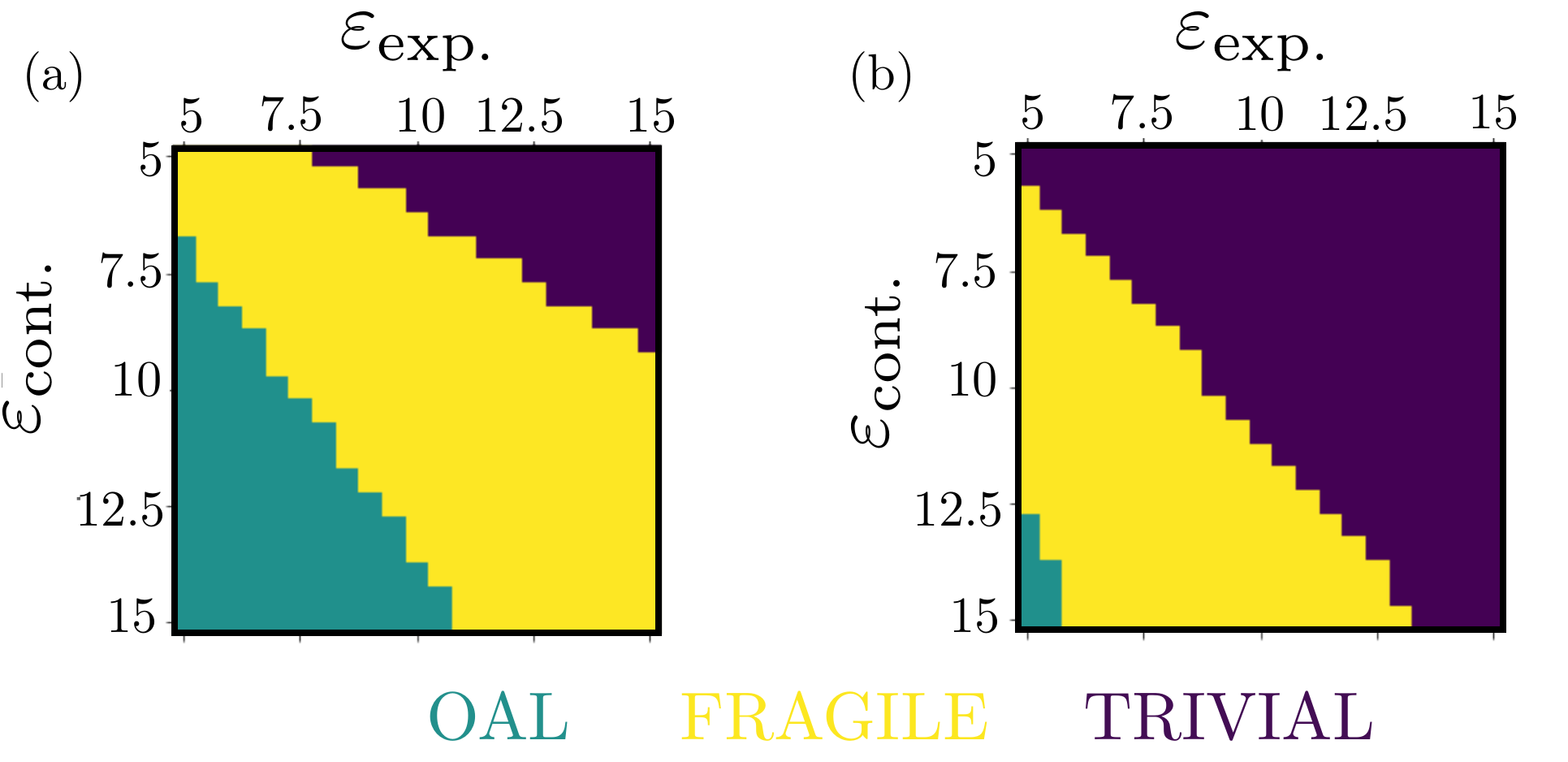}    
    \end{flushright}
    \caption{\label{figure_12} Phase diagram for the NBHL as a function of the different values of the dielectric constant for the contracted $\varepsilon_\mathrm{cont.}$ and expanded $\varepsilon_\mathrm{exp.}$ lattices: (a) $r_\mathrm{cont.}=0.1$ and $r_\mathrm{exp.}=0.05$ (b) $r_\mathrm{cont.}=0.1$ and $r_\mathrm{exp.}=0.075$.}
\end{figure}

We conclude by addressing the robustness of the results presented for the three possible configurations of the nested breathing honeycomb lattice. We presented results for specific values of the dielectric constants and radii for the dielectric rods in the contracted and expanded positions. In Fig.~\ref{figure_12} we present the phase diagram of the NBHL obtained for fixed radii of the rods in contracted and expanded positions and varying the value of their dielectric constants.
%
%
%%%%%%%%%%%
%%%%%%%%%%%
%
%
For both choices of radii, we observe that it is possible to obtain in the phase diagram all three phases including the fragile topological one. In this respect, the PhC based on NBHL is a new platform for investigating the fragile topological phase. The phase diagram has been obtained by using the tools of band representation similar to Ref.~\cite{FragileLight}.

\section{Acknowledgements}
We acknowledge useful discussions with Alessandro De Martino, Duy Hoang Minh Nguyen, and Ivo Souza. The work of M.A.J.H. and D.B. is supported from Ministerio de Ciencia e Innovaci\'on (MICINN) through Project No.~PID2020-120614GB-I00, and by the Transnational Common Laboratory $Quantum-ChemPhys$ (D.B.). Additionally, M.B.P., G.G., A.G.E., M.G.V. and D.B. acknowledge Programa Red guipuzcoana de Ciencia, Tecnolog\'ia e Innovaci\'on 2021, Grant nr. 2021-CIEN-000070-01, Gipuzkoa Next, and the funding from the Basque Government's IKUR initiative on Quantum technologies (Department of Education). M.G.V. acknowledges the Spanish Ministerio de Ciencia e Innovacion (grant PID2019-109905GB-C21).
A.G.E. and M.B.P. acknowledge support from the Spanish Ministerio de Ciencia e Innovaci\'on (PID2019-109905GA-C2) and from Eusko Jaurlaritza (IT1164-19 and KK-2021/00082). 
P.A.H. acknowledges funding from Funda\c c\~ao para a Ci\^encia e a Tecnologia and Instituto de Telecomunica\c c\~oes under project UIDB/50008/2020 and the CEEC Individual program with reference CEECIND/02947/2020. H. A. acknowledges the Purdue University Startup fund. The work of B.~B. is supported by the Air Force Office of Scientific Research under award number FA9550-21-1-0131.

\appendix
\section{Dicrete Wilson loop}\label{app_one}
In this Appendix, we show how to evaluate the WL spectrum for a lattice system that is sampled on a finite set of momenta in the first BZ.

The WL operator is computed via a path-ordered integral along one of the reciprocal directions. Because of the periodic boundary conditions, this path is closed.  
Arbitrary phases added by numerical diagonalization procedures will be different at each $\mathbf{k}$-point, and since the WL operator is gauge covariant, it will depend on those phases (though its spectrum will not). Therefore, we need to ensure periodicity to remove any arbitrary phase to get good convergence with the density of the BZ grid. A procedure for fixing this problem is given in Ref.~\cite{tutorial}, which we review here.

We start by considering the most simple case of a single band isolated from the rest by energy gaps. We characterize the discretized the path along the reciprocal lattice vectors ($\mathbf{k}_\alpha$,$\mathbf{k}_\beta$) with the  points $(k_i,k_j), j=1,\dots N$ with $k_i$ fixed and we calculate the discrete Wilson loop as,
%
%
%%%%%%%%%%%%%%%
\begin{equation}\label{Wilson:one:discret}
    \mathcal{W}(k_i)= - \mathrm{Im}   \left[ \log \left(  \prod_{k_j}\langle u_{(k_i,k_j)}| u_{(k_i,k_{j+1})}\rangle \right)\right]\, .
\end{equation}
%%%%%%%%%%%%%%%
%
%
This formula indicates that the Wilson loop can be computed for each $k_i$ along the $k_j$ direction by taking the phase of the final product of the overlap of consecutive evaluations of the periodic part of the Bloch wave function over the mesh of reciprocal space points.

On the other hand, for the case of an isolated composite group of bands, we replace the scalar product by overlap matrices $S^{mn}_{(k_i,k_j)\rightarrow(k_i,k_{j+1})}$ 
%
%
%%%%%%%%%%%%%%%
\begin{equation}
     \mathcal{W}_{mn}(k_i)= - \mathrm{Im}\left[ \log \left( \prod_{k_j} \mathcal{S}^{mn}_{(k_i,k_j)\rightarrow(k_i,k_{j+1})}\ \right)\right]\, .  
\end{equation}
%%%%%%%%%%%%%%%
%
%
The overlap matrix, $\mathcal{S}$, between $(k_i,k_j)$ and $(k_i,k_{j+1})$ can be expressed as, 
%
%
%%%%%%%%%%%%%%%
\begin{equation*}
\hspace{-1.5cm} 
\mathcal{S}_{(k_i,k_j)\rightarrow (k_i,k_{j+1})}^{mn}=
\left(
\begin{array}{c c c}
%  line 1
  \langle u^{1}_{(k_i,k_{j})}|u^{1}_{(k_i,k_{j+1})}\rangle &\ldots&
  \langle u^1_{(k_i,k_{j})}|u^{n}_{(k_i,k_{j+1})}\rangle\\
% line 2
   \vdots &\ddots & \vdots\\
% line 3
    \langle u^{n}_{(k_i,k_{j})}|u^{1}_{(k_i,k_{j+1})}\rangle &
    \ldots & \langle u^{n}_{(k_i,k_{j})}|u^{n}_{(k_i,k_{j+1})}\rangle
     
\end{array}
\right)%}
\end{equation*}
%%%%%%%%%%%%%%%
%
%
where the superscript $\ell$ of $u^\ell_{\mathbf{k}}$ indicates the band index of the Bloch mode. 
%In this case, we construct the overlap matrices between consecutive points along the lines in the $k_j$ direction. 
Then we multiply the overlap matrices for each pair of points, and finally, the resulting matrix is diagonalized. The phases of its eigenvalues encode information about the position of the Wannier centers in real space. In Ref.~\cite{tutorial}, it is explained in detail how to use the overlap matrix for PhCs.
In two dimensions, one of the momenta $k_j$ defines the integration variable of the closed path $\ell$ in Eq.~(\ref{eq:WL}), while the other momentum $k_i$ is a free parameter characterizing the Wilson loop.

\section{Poynting vector and EM energy density}\label{app_two}
In this appendix we summarize the Poynting vector formalism, a well-known tool for calculating the power flow of EM radiation, and show its relationship to the EM energy density as used throughout this text to normalize the fields. 

From the Lorentz force law it is straightforward to see that the instantaneous power density is determined as
%
%
%%%%%%%%%%%%%%
\begin{equation}%~\label{eq:inst. power}
    p(t) = \boldsymbol{\mathcal{E}}(t) \cdot \boldsymbol{\mathcal{J}}(t) \, ,
\end{equation}
%%%%%%%%%%%%%%
%
%
where $\boldsymbol{\mathcal{E}},\boldsymbol{\mathcal{J}}$ are the electric field and current density vector, respectively.

When re-written in terms of the electromagnetic fields only, the instantaneous power density gets the following form
%
%
%%%%%%%%%%%%%%
\begin{equation}%~\label{eq:diff. form Poynting}
    p(t) = - \nabla \cdot \left(\boldsymbol{\mathcal{E}}(t) \times \boldsymbol{\mathcal{H}}(t) \right) - \boldsymbol{\mathcal{H}}(t) \cdot \frac{\partial \boldsymbol{\mathcal{B}}(t)}{\partial t} - \boldsymbol{\mathcal{E}}(t) \cdot \frac{\partial \boldsymbol{\mathcal{D}}(t)}{\partial t} \, .
\end{equation}
%%%%%%%%%%%%%%
%
%
This already shows the differential form of the energy balance flow: the LHS is the instantaneous power, while the first term on the RHS is the differential form of the Poynting vector, \emph{i.e.} $\boldsymbol{\mathcal{S}}(t) = \boldsymbol{\mathcal{E}}(t) \times \boldsymbol{\mathcal{H}}(t) $, corresponding to the power flux, and the second and third term express the change rate of the stored energy in the magnetic and electric fields, respectively.

In the absence of source, \emph{i.e.}, $\boldsymbol{\mathcal{J}}(t) = 0$, the above equation will be simplified to the well-known Poynting theorem relating the outgoing power from a closed surface to the changing rate of the stored electromagnetic energy within the volume as
%
%
%%%%%%%%%%%%%%
\begin{equation}
    \oint \boldsymbol{\mathcal{S}}(t) \cdot d\boldsymbol{s} = - \int dv \left (\boldsymbol{\mathcal{H}}(t) \cdot \frac{\partial \boldsymbol{\mathcal{B}}(t)}{\partial t} +  \boldsymbol{\mathcal{E}}(t) \cdot \frac{\partial \boldsymbol{\mathcal{D}}(t)}{\partial t} \right) \, .
\end{equation}
%%%%%%%%%%%%%%
%
%
The stored energy on the right-hand side, for a time-harmonic electromagnetic field in a non-magnetic and non-dispersive medium can be simplified to
%
%
%%%%%%%%%%%%%%%%
\begin{equation}
    \braket{\epsilon} = \int dv \left(\frac{\mu_0}{4} |\mathbf{H}(r)|^2 + \frac{\varepsilon_0 \varepsilon(r)}{4} |\mathbf{E}(r)|^2 \right)\, ,
\end{equation}
%%%%%%%%%%%%%%%%
%
%
where $\mathbf{H}(r)$, $\mathbf{E}(r)$ are the magnetic and electric field phasors and the first and second terms in the kernel correspond to the time-averaged magnetic and electric energy densities, respectively.

The above form can be simplified further in terms of the electric field only, using the Maxwell's equations relating the magnetic field phasor to the electric field as
%
%
%%%%%%%%%%%%%%%%
\begin{eqnarray}
    \braket{\epsilon_\mathrm{m}} & = \frac{\mu_0}{4} \int_\mathrm{UC} dv |\boldsymbol{H}(r)|^2 = \frac{1}{4\omega^2 \mu_0} \int_\mathrm{UC} dv \left(\boldsymbol\nabla \times \mathbf{E} \right) \cdot \left(\boldsymbol\nabla \times \mathbf{E}^*\right) \\ \nonumber
    & = \frac{1}{4\omega^2 \mu_0} \oint \left(\mathbf{E} \times \boldsymbol\nabla \times \mathbf{E}^* \right) \cdot d\boldsymbol{s} + \frac{1}{4\omega^2 \mu_0} \int_\mathrm{UC} dv \left(\mathbf{E} \cdot \boldsymbol\nabla\times \boldsymbol\nabla \times \mathbf{E}^* \right) \\ \nonumber
    & = \frac{1}{4\omega^2\mu_0} \int_\mathrm{UC} -\mathrm{i}\omega \mu_0 \left(\mathbf{E} \cdot \boldsymbol\nabla  \times \mathbf{H}^*\right)dv = \frac{1}{4} \int_\mathrm{UC} \varepsilon_0 \varepsilon(r) |\mathbf{E}(r)|^2dv \, ,
\end{eqnarray}
%%%%%%%%%%%%%%%%
%
%
where the closed-surface integral in the second line vanishes due to the periodic boundary conditions of the unit cell.

With this the final form of the stored energy can be expressed in terms of the electric field solely as
%
%
%%%%%%%%%%%%%%
\begin{eqnarray}
    \braket{\epsilon} =\frac{\varepsilon_0}{2} \int dv \,  \varepsilon(r) |\mathbf{E}(r)|^2 \, ,
\end{eqnarray}
%%%%%%%%%%%%%%
%
%
leading to the following definition of the EM energy density as $n(\mathbf{r}) = \frac{1}{2} \varepsilon_0 \varepsilon(r) |\mathbf{E}(r)|^2$ as it is customary for the field quantization and has been introduced in the main text.

Finally, we want to point out that the equivalence of the stored electric and magnetic energy densities, \emph{i.e.}, $\braket{\epsilon_\mathrm{m}} = \braket{\epsilon_\mathrm{e}}$, as derived above is the direct consequence of the \emph{Virial} theorem.

\section*{References}
\bibliographystyle{iopart-num}
\bibliography{references}
\end{document}